\pdfoutput=1

\documentclass[10pt,letterpaper]{article}
\usepackage[top=0.85in,left=1.00in,footskip=0.75in]{geometry}
\usepackage[utf8]{inputenc}
\usepackage{amsmath,amssymb}
\usepackage{cite}

\usepackage{color}
\usepackage[dvipsnames]{xcolor}
\definecolor{cuteBlue}{rgb}{0.258, 0.387, 0.574}
\definecolor{cuteGreen}{rgb}{0, 0.3, 0}

\usepackage[normalem]{ulem}

\usepackage{nameref}
\usepackage[colorlinks=true, urlcolor=cuteBlue, citecolor=cuteGreen, linkcolor=black]{hyperref}

\usepackage{microtype}
\DisableLigatures[f]{encoding = *, family = *}

\usepackage{rotating}
\usepackage{pdflscape}

\usepackage{float}

\usepackage[aboveskip=1pt,labelfont=bf,labelsep=period,justification=raggedright,singlelinecheck=off]{caption}

\makeatletter
\renewcommand{\@biblabel}[1]{\quad#1.}
\makeatother

\usepackage{lineno}
\usepackage{xparse}

\DeclareDocumentCommand \eref{oooo} {\IfNoValueTF{#2}{Eq.~(\ref{#1})}{\IfNoValueTF{#3}{Eqs.~(\ref{#1}) and (\ref{#2})}{\IfNoValueTF{#4}{Eqs.~(\ref{#1})-(\ref{#3})}{Eqs.~(\ref{#1})-(\ref{#4})}}}}

\DeclareDocumentCommand \fref{ooo} {\IfNoValueTF{#2}{Fig.~\ref{#1}}{\IfNoValueTF{#3}{Figs.~\ref{#1} and \ref{#2}}{Figs.~\ref{#1}-\ref{#3}}}}

\newcommand{\letter}[1]{#1} % For main text. Ex: As shown in Fig. 12A
\newcommand{\letterParen}[1]{(#1)} % For captions. Ex: (A) Low limit and (B) high limit

\newcommand \foldchange{\operatorname{fold-change}}

\def\true{True}

\def\showAll{False}
\def\showText{True}

\ifx\showAll\true
\else
	\ifx\showText\true
	\else
	\fi
\fi

\begin{document}
	
	% Main text files
	\begin{flushleft}

{\Large
		\textbf\newline{Tuning transcriptional regulation through signaling: A predictive theory of allosteric induction}}
	\newline
	% Insert Author names, affiliations and corresponding author email.
	\\
	\textbf{Manuel Razo-Mejia$^{1,\dagger}$, Stephanie L. Barnes$^{1,\dagger}$, Nathan M. Belliveau$^{1,\dagger}$, Griffin Chure$^{1,\dagger}$, Tal Einav$^{2,\dagger}$, Mitchell Lewis$^{3}$, Rob Phillips$^{1,4,*}$}
	\\
	$^1$Division of Biology and Biological Engineering, California Institute of Technology,
	Pasadena, CA, United States;
	$^2$Department of Physics, California Institute of Technology,
	Pasadena, CA, United States;
	$^3$Department of Biochemistry and Biophysics, University of Pennsylvania  School of Medicine, Philadelphia, PA, United  States;
	$^4$Department of Applied Physics, California Institute of Technology,
	Pasadena, CA, United States
    \\
	$^\dagger$ contributed equally
	\\
	* phillips@pboc.caltech.edu
    
\end{flushleft}

%%%%%%%%%%%%%%%%%%%%%%%%%%%%%%%%%%%%%%%%%%%%%%%%%%%%%%%%%%%%%%%%%%%%%%%%%%%
\section*{Abstract}

Allosteric regulation is found across all domains of life, yet we still lack
simple, predictive theories that directly link the experimentally tunable
parameters of a system to its input-output response. To that end, we present a
general theory of allosteric transcriptional regulation using the
Monod-Wyman-Changeux model. We rigorously test this model using the ubiquitous
simple repression motif in bacteria by first predicting the behavior of strains
that span a large range of repressor copy numbers and DNA binding strengths and
then constructing and measuring their response. Our model not only accurately
captures the induction profiles of these strains but also enables us to derive
analytic expressions for key properties such as the dynamic range and
$[EC_{50}]$. Finally, we derive an expression for the free energy of allosteric
repressors which enables us to collapse our experimental data onto a single
master curve that captures the diverse phenomenology of the induction profiles.

\section*{Introduction}

Understanding how organisms sense and respond to changes in their environment
has long been a central theme of biological inquiry. At the cellular level,
this interaction is mediated by a diverse collection of molecular signaling
pathways. A pervasive mechanism of signaling in these pathways is allosteric regulation, in
which the binding of a ligand induces a conformational change in some target
molecule, triggering a signaling cascade \cite{Lindsley2006}. One of the most
important examples of such signaling is offered by transcriptional regulation,
where a transcription factor's propensity to bind to DNA will be altered upon
binding to an allosteric effector.

Despite the overarching importance of this mode of signaling, a quantitative
understanding of the molecular interactions between extracellular inputs and
gene expression remains poorly explored. Attempts to reconcile theoretical
models and experiments have often been focused on fitting data retrospectively
after experiments have been conducted \cite{Kuhlman2007, Daber2009}. Further,
many treatments of induction are strictly phenomenological, electing to treat
induction curves individually either using Hill functions or as ratios of
polynomials without acknowledging that allosteric proteins have distinct
conformational states depending upon whether an effector molecule is bound to
them or not \cite{Setty2003,Poelwijk2011,Vilar2013,Rogers2015,Rohlhill2017}.
These fits are made in experimental conditions in which there is great
uncertainty about the copy number of both the transcription factor and the regulated locus, meaning that
the underlying minimal set of parameters cannot be pinned down unequivocally.
This leaves little prospect for predicting or understanding what molecular
properties determine key phenotypic parameters such as leakiness, dynamic range,
$[EC_{50}]$, and the effective Hill coefficient as discussed in
Refs.~\cite{Martins2011, Marzen2013} and illustrated in
\fref[figInductionCorepressionPhenotypicProperties]. Our goal was to use a minimal
Monod-Wyman-Changeux (MWC) model of transcription factor induction in
conjunction with a corresponding thermodynamic model of repression to test
whether such a simple model is capable of predicting how the induction process
changes over broad swathes of regulatory parameter space. While some treatments
of induction have used MWC models to predict transcriptional outputs
\cite{Daber2009, Daber2011a, Sochor2014}, these often require multi-parameter
fitting which gives rise to issues of parameter degeneracy (see Appendix
A) and may include effective parameters that have tenuous biological meaning. In contrast, our objective was to use the MWC model to
make parameter-free predictions about how the induction response will be altered
when transcription factor copy number and operator strength are systematically
varied.

We test our model in the context of the simple repression motif -- a widespread
bacterial genetic regulatory architecture in which binding of a transcription
factor occludes binding of an RNA polymerase thereby inhibiting transcription
initiation. A recent survey of different regulatory architectures within the
{\it E. coli} genome revealed that more than 100 genes are characterized by the
simple repression motif, making it a common and physiologically relevant
architecture \cite{Rydenfelt2014}. Building upon previous work
\cite{Garcia2011,Brewster2014,Weinert2014}, we present a statistical mechanical
rendering of allostery in the context of induction and corepression, shown
schematically in \fref[figInductionCorepressionPhenotypicProperties]\letter{A},
and use this model as the basis of parameter-free predictions which we then
probe experimentally. Specifically, we model the allosteric response of
transcriptional repressors using the MWC model, which stipulates that an
allosteric protein fluctuates between two distinct conformations -- an active
and inactive state -- in thermodynamic equilibrium \cite{MONOD1965}. In the
context of induction, effector binding increases the probability that a
repressor will be in the inactive state, weakening its ability to bind to the
promoter and resulting in increased expression. The framework presented here
provides considerable insight beyond that of simply fitting a sigmoidal curve to
inducer titration data. We aim to explain and predict the relevant biologically
important parameters of an induction profile, such as characterizing the
midpoint and steepness of its response as well as the limits of minimum and
maximum expression as shown in
\fref[figInductionCorepressionPhenotypicProperties]\letter{B}. By combining this
MWC treatment of induction with a thermodynamic model of transcriptional
regulation (\fref[fig_polymerase_repressor_states]), we create a general
quantitative model of allosteric transcriptional regulation that is applicable
to a wide range of regulatory architectures such as activation, corepression,
and various combinations thereof, extending our quantitative understanding of these schemes \cite{Bintu2005} to include signaling.

To demonstrate the predictive power of our theoretical formulation across a wide range of both operator strengths and repressor copy
numbers, we design an \textit{E. coli} genetic construct in which the binding probability of a
repressor regulates gene expression of a fluorescent reporter. Using components
from the well-characterized \textit{lac} system in \textit{E. coli}, we first
quantify the three parameters associated with the induction of the repressor,
namely, the binding affinity of the active and inactive repressor to the inducer
and the free energy difference between the active and inactive repressor states.
We determine these parameters by fitting to measurements of the fold-change in
gene expression as a function of inducer concentration for a circuit with known
repressor copy number and repressor-operator binding energy. We note that all
other parameters that appear in the thermodynamic model are used without change
from a suite of earlier experiments which quantify fold-change in a range of
regulatory scenarios \cite{Garcia2011, Garcia2011B, Brewster2012,
	Boedicker2013a, Boedicker2013b, Brewster2014}. With these estimated allosteric
parameters in hand, we make accurate, parameter-free predictions of the
induction response for many other combinations of repressor copy number and
binding energy. This goes well beyond previous treatments of the induction
phenomenon and shows that one extremely compact set of parameters can be applied
self-consistently and predictively to vastly different regulatory situations
including simple repression on chromosome, cases in which decoy binding sites
for repressor are put on plasmids, cases in which multiple genes compete for the
same regulatory machinery, cases involving multiple binding sites for repressor leading to DNA looping, and the induction experiments described
here. The broad reach of this minimal parameter set is highlighted in
\fref[fig_previous_work].

Rather than viewing the behavior of each circuit as giving rise to its own
unique input-output response, the formulation of the MWC model presented here
provides a means to characterize these seemingly diverse behaviors using a
single unified framework governed by a small set of parameters, applicable even
to mutant repressors in much the same way that earlier work showed how mutants
in quorum sensing and chemotaxis receptors could be understood within a minimal
MWC-based model \cite{Swem2008, Keymer2006}. Another insight that emerges from
our theoretical treatment is how a subset in the parameter space of repressor
copy number, operator binding site strength, and inducer concentration can all
yield the same level of gene expression. Our application of the MWC model allows
us to understand these degeneracies in parameter space through an expression for
the free energy of repressor binding, a nonlinear combination of physical
parameters which determines the system's mean response and is the fundamental
quantity that dictates the phenotypic cellular response to a signal.

\begin{figure}[h!]
	\centering \includegraphics{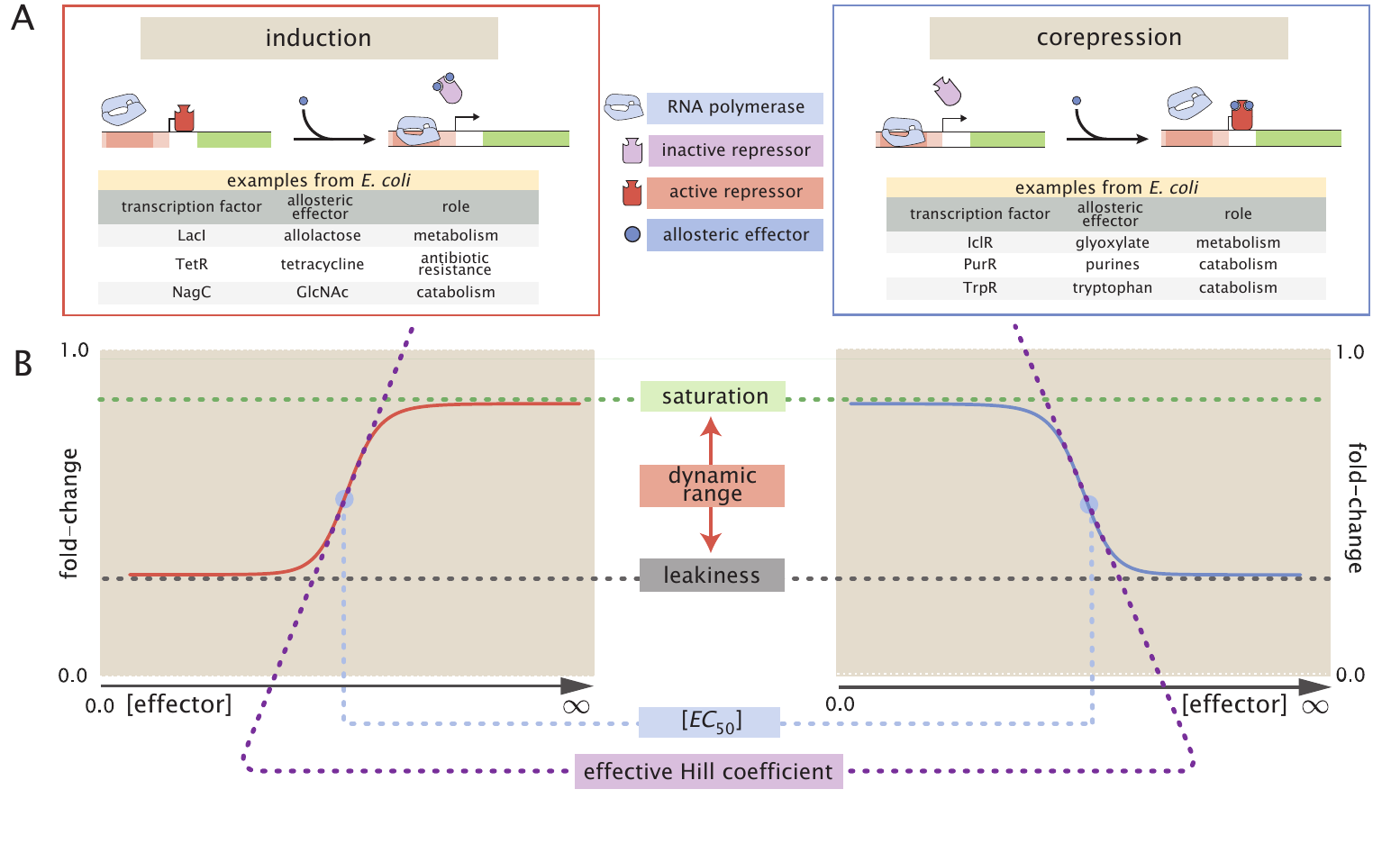}

	\caption{{\bf Transcription regulation architectures involving an allosteric
		repressor.} \letterParen{A} We consider a promoter regulated solely by an
	allosteric repressor. When bound, the repressor prevents RNAP from binding and
	initiating transcription. Induction is characterized by the addition of an
	effector which binds to the repressor and stabilizes the inactive state
	(defined as the state which has a low affinity for DNA), thereby increasing
	gene expression. In corepression, the effector stabilizes the repressor's
	active state and thus further reduces gene expression. We list several
	characterized examples of induction and corepression that support different
	physiological roles in {\it E. coli} \cite{Huang2011,Li2014}. \letterParen{B} A
	schematic regulatory response of the two architectures shown in Panel
	\letter{A} plotting the fold-change in gene expression as a function of
	effector concentration, where fold-change is defined as the ratio of gene
	expression in the presence versus the absence of repressor. We consider the
	following key phenotypic properties that describe each response curve: the
	minimum response (leakiness), the maximum response (saturation), the difference
	between the maximum and minimum response (dynamic range), the concentration of
	ligand which generates a fold-change halfway between the minimal and maximal
	response ($[EC_{50}]$), and the log-log slope at the midpoint of the response
	(effective Hill coefficient).}
\label{figInductionCorepressionPhenotypicProperties}
\end{figure}

	%%%%%%%%%%%%%%%%%%%%%%%%%%%%%%%%%%%%%%%%%%%%%%%%%%%%%%%%%%%%%%%%%%%%%%%%%%%
\section*{Results}

\subsection*{Characterizing Transcription Factor Induction using the Monod-Wyman-Changeux (MWC) Model}

We begin by considering the induction of a simple repression genetic
architecture, in which the binding of a transcriptional repressor occludes the
binding of RNA polymerase (RNAP) to the DNA \cite{Ackers1982,Buchler2003}. When
an effector (hereafter referred to as an ``inducer" for the case of induction)
binds to the repressor, it shifts the repressor's allosteric equilibrium towards
the inactive state as specified by the MWC model \cite{MONOD1965}. This causes
the repressor to bind more weakly to the operator, which increases gene
expression. Simple repression motifs in the absence of inducer have been
previously characterized by an equilibrium model where the probability of each
state of repressor and RNAP promoter occupancy is dictated by the Boltzmann
distribution \cite{Buchler2003, Vilar2003, Bintu2005a, Garcia2011, Brewster2014,
	Ackers1982} (we note that non-equilibrium models of simple repression have been
shown to have the same functional form that we derive below
\cite{Phillips2015a}). We extend these models to consider the role of allostery
by accounting for the equilibrium state of the repressor through the MWC model
as follows.

Consider a cell with copy number \(P\) of RNAP and \(R\) repressors. Our model
assumes that the repressor can exist in two conformational states. \(R_A\)
repressors will be in the active state (the favored state when the repressor is
not bound to an inducer; in this state the repressor binds tightly to the DNA)
and the remaining \(R_I\) repressors will be in the inactive state (the
predominant state when repressor is bound to an inducer; in this state the
repressor binds weakly to the DNA) such that \(R_A+R_I=R\). Repressors fluctuate
between these two conformations in thermodynamic equilibrium \cite{MONOD1965}.

Thermodynamic models of gene expression begin by enumerating all possible states
of the promoter and their corresponding statistical weights. As shown in
\fref[fig_polymerase_repressor_states]\letter{A}, the promoter can either be
empty, occupied by RNAP, or occupied by either an active or inactive repressor.
We assign the repressor a different DNA binding affinity in the active and
inactive state. In addition to the specific binding sites at the promoter, we
assume that there are $N_{NS}$ non-specific binding sites elsewhere (i.e. on
parts of the genome outside the simple repression architecture) where the RNAP
or the repressor can bind. All specific binding energies are measured relative
to the average non-specific binding energy. Our model explicitly ignores the
complexity of the distribution of non-specific binding affinities in the genome,
and makes the assumption that a single parameter can capture the energy
difference between our binding site of interest and the average site in the
reservoir. Thus, \(\Delta\varepsilon_{P}\) represents the energy difference
between the specific and non-specific binding for RNAP to the DNA. Likewise,
\(\Delta\varepsilon_{RA}\) and \(\Delta\varepsilon_{RI}\) represent the
difference in specific and non-specific binding energies for repressor in the
active or inactive state, respectively.

\begin{figure}
	\centering \includegraphics{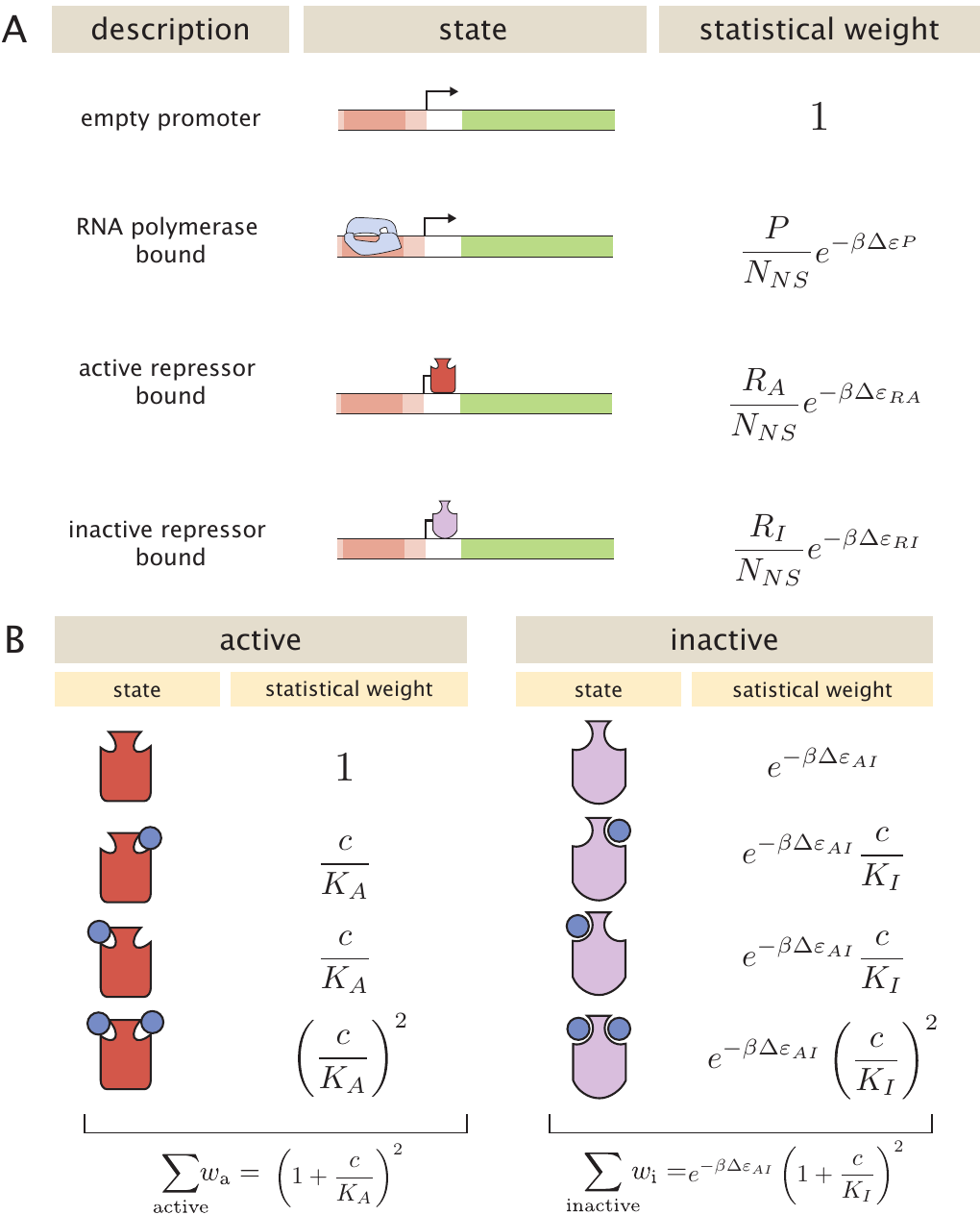}

	\caption{\textbf{States and weights for the simple repression motif.}
	\letterParen{A} RNAP (light blue) and a repressor compete for binding to a
	promoter of interest. There are $R_A$ repressors in the active state (red) and
	$R_I$ repressors in the inactive state (purple). The difference in energy
	between a repressor bound to the promoter of interest versus another
	non-specific site elsewhere on the DNA equals $\Delta\varepsilon_{RA}$ in the
	active state and $\Delta\varepsilon_{RI}$ in the inactive state; the $P$ RNAP
	have a corresponding energy difference $\Delta\varepsilon_{P}$ relative to
	non-specific binding on the DNA. $N_{NS}$ represents the number of non-specific
	binding sites for both RNAP and repressor. \letterParen{B} A repressor has an
	active conformation (red, left column) and an inactive conformation (purple, right
	column), with the energy difference between these two states given by $\Delta
	\varepsilon_{AI}$. The inducer (blue circle) at concentration $c$ is capable of
	binding to the repressor with dissociation constants $K_A$ in the active state
	and $K_I$ in the inactive state. The eight states for a dimer with $n=2$
	inducer binding sites are shown along with the sums of the active and inactive states.} \label{fig_polymerase_repressor_states}
\end{figure}

Thermodynamic models of transcription \cite{Daber2011a,Kuhlman2007, Weinert2014, Buchler2003,
	Vilar2003, Bintu2005a, Bintu2005, Garcia2011, Brewster2014, Ackers1982} posit that gene
expression is proportional to the probability that the RNAP is bound to the
promoter $p_{\text{bound}}$, which is given by
\begin{equation}\label{eq_p_bound_definition}
p_\text{bound}=\frac{\frac{P}{N_{NS}}e^{-\beta \Delta\varepsilon_{P}}}{1+\frac{R_A}{N_{NS}}e^{-\beta \Delta\varepsilon_{RA}}+\frac{R_I}{N_{NS}}e^{-\beta \Delta\varepsilon_{RI}}+\frac{P}{N_{NS}}e^{-\beta\Delta\varepsilon_{P}}},
\end{equation}
with $\beta = \frac{1}{k_BT}$ where $k_B$ is the Boltzmann constant and $T$ is
the temperature of the system. As $k_BT$ is the natural unit of energy at the
molecular length scale, we treat the products $\beta \Delta\varepsilon_{j}$ as
single parameters within our model. Measuring $p_{\text{bound}}$ directly is
fraught with experimental difficulties, as determining the exact proportionality between expression and $p_{\text{bound}}$ is not straightforward. Instead, we measure the
fold-change in gene expression due to the presence of the repressor. We define
fold-change as the ratio of gene expression in the presence of repressor
relative to expression in the absence of repressor (i.e. constitutive expression), namely,
\begin{equation}\label{eq_fold_change_definition}
\foldchange \equiv \frac{p_\text{bound}(R > 0)}{p_\text{bound}(R = 0)}.
\end{equation}
We can simplify this expression using two well-justified approximations: (1)
$\frac{P}{N_{NS}}e^{-\beta\Delta\varepsilon_{P}}\ll 1$ implying that the RNAP
binds weakly to the promoter ($N_{NS} = 4.6 \times 10^6$, $P \approx 10^3$
\cite{Klumpp2008} , $\Delta\varepsilon_{P} \approx -2 \,\, \text{to} \, -5~k_B
T$ \cite{Brewster2012}, so that $\frac{P}{N_{NS}}e^{-\beta\Delta\varepsilon_{P}}
\approx 0.01$) and (2) $\frac{R_I}{N_{NS}}e^{-\beta \Delta\varepsilon_{RI}} \ll
1 + \frac{R_A}{N_{NS}} e^{-\beta\Delta\varepsilon_{RA}}$ which reflects our
assumption that the inactive repressor binds weakly to the promoter of interest.
Using these approximations, the fold-change reduces to the form
\begin{equation}\label{eq_fold_change_approx}
\foldchange \approx \left(1+\frac{R_A}{N_{NS}}e^{-\beta \Delta\varepsilon_{RA}}\right)^{-1} \equiv \left( 1+p_A(c) \frac{R}{N_{NS}}e^{-\beta
	\Delta\varepsilon_{RA}} \right)^{-1},
\end{equation}
where in the last step we have introduced the fraction $p_A(c)$ of repressors in
the active state given a concentration $c$ of inducer, which is defined as
$R_A(c)=p_A(c) R$. Since inducer binding shifts the repressors from the active
to the inactive state, $p_A(c)$ is a decreasing function of $c$
\cite{Marzen2013}.

We compute the probability $p_A(c)$ that a repressor with $n$ inducer binding
sites will be active using the MWC model. After first enumerating all possible
configurations of a repressor bound to inducer (see
\fref[fig_polymerase_repressor_states]\letter{B}), $p_A(c)$ is given by the sum
of the weights of the active states divided by the sum of the weights of all
possible states, namely,
\begin{equation}\label{eq_p_active}
p_A(c)=\frac{\left(1+\frac{c}{K_A}\right)^n}{\left(1+\frac{c}{K_A}\right)^n+e^{-\beta \Delta \varepsilon_{AI} }\left(1+\frac{c}{K_I}\right)^n},
\end{equation}
where $K_A$ and $K_I$ represent the dissociation constant between the inducer
and repressor in the active and inactive states, respectively, and $\Delta
\varepsilon_{AI} = \varepsilon_{I} - \varepsilon_{A}$ stands for the free energy
difference between a repressor in the inactive and active state (the quantity
$e^{-\Delta \varepsilon_{AI}}$ is sometimes denoted by $L$ \cite{Marzen2013, MONOD1965} or
$K_{\text{RR}*}$ \cite{Daber2011a}). A repressor which favors the active state
in the absence of inducer ($\Delta \varepsilon_{AI} > 0$) will be driven towards
the inactive state upon inducer binding when $K_I < K_A$. The specific case of a
repressor dimer with $n=2$ inducer binding sites is shown in
\fref[fig_polymerase_repressor_states]\letter{B}.

Substituting $p_A(c)$ from \eref[eq_p_active] into \eref[eq_fold_change_approx]
yields the general formula for induction of a simple repression regulatory
architecture, namely,
\begin{equation}\label{eq_fold_change_full}
\foldchange = \left(
1+\frac{\left(1+\frac{c}{K_A}\right)^n}{\left(1+\frac{c}{K_A}\right)^n+e^{-\beta \Delta \varepsilon_{AI} }\left(1+\frac{c}{K_I}\right)^n}\frac{R}{N_{NS}}e^{-\beta \Delta\varepsilon_{RA}} \right)^{-1}.
\end{equation}
While we have used the specific case of simple repression with induction to
craft this model, we reiterate that the exact same mathematics describe the case
of corepression in which binding of an allosteric effector stabilizes the active
state of the repressor and decreases gene expression (see
\fref[figInductionCorepressionPhenotypicProperties]\letter{B}). Interestingly,
we shift from induction (governed by $K_I < K_A$) to corepression ($K_I > K_A$)
as the ligand transitions from preferentially binding to the inactive repressor
state to stabilizing the active state. Furthermore, this general approach can be
used to describe a variety of other motifs such as activation, multiple
repressor binding sites, and combinations of activator and repressor binding
sites \cite{Brewster2014, Weinert2014, Bintu2005}.

This key formula presented in \eref[eq_fold_change_full] enables us to make
precise quantitative statements about induction profiles. Motivated by the broad
range of predictions implied by this equation, we designed a series of
experiments using the \textit{lac} system in \textit{E. coli} to tune the
control parameters for a simple repression genetic circuit. As discussed in
\fref[fig_previous_work], previous studies from our lab have provided us with
well-characterized values for many of the parameters in our experimental system,
leaving only the values of the the MWC parameters ($K_A$, $K_I$, and $\Delta
\varepsilon_{AI}$) to be determined. We note that while previous studies have
obtained values for $K_A$, $K_I$, and $L=e^{-\beta \Delta \varepsilon_{AI}}$
\cite{Daber2011a, OGorman1980}, they were either based upon clever biochemical
experiments or \textit{in vivo} conditions involving poorly characterized
transcription factor copy numbers and gene copy numbers. These differences
relative to our experimental conditions and fitting techniques led us to believe
that it was important to perform our own analysis of these parameters. Indeed,
after inferring these three MWC parameters (see Appendix A for details regarding the inference of $\Delta \varepsilon_{AI}$, which was fitted separately from $K_A$ and $K_I$), we were able to predict the
input/output response of the system under a broad range of experimental
conditions. For example, this framework can predict the response of the system
at different repressor copy numbers $R$, repressor-operator affinities
$\Delta\varepsilon_{RA}$, inducer concentrations $c$, and gene copy numbers (see
Appendix B).

	%%%%%%%%%%%%%%%%%%%%%%%%%%%%%%%%%%%%%%%%%%%%%%%%%%%%%%%%%%%%%%%%%%%%%%%%%%%
\subsection*{Experimental Design}

To test this model of allostery, we build off of a collection of work that has
developed both a quantitative understanding of and experimental control over the
simple repression motif. As shown in \fref[fig_previous_work], earlier work from
our laboratory used \textit{E. coli} constructs based on components of the
\textit{lac} system to demonstrate how the Lac repressor (LacI) copy number $R$
and operator binding energy $\Delta\varepsilon_{RA}$ affect gene expression in
the absence of inducer \cite{Garcia2011}. Rydenfelt \textit{et
	al.}~\cite{Rydenfelt2014B} extended the theory used in that work to the case of
multiple promoters competing for a given transcription factor, which was
demonstrated experimentally by Brewster \textit{et al.}~\cite{Brewster2014}, who
modified this system to consider expression from multiple-copy plasmids as well
as the presence of competing repressor binding sites. Although the current work
focuses on systems with a single site of repression, in Appendix A we
utilize data from Brewster \textit{et al.}~\cite{Brewster2014} to characterize
the allosteric free energy difference $\Delta\varepsilon_{AI}$ between the
repressor's active and inactive states. With this parameter in hand, the present
work considers the effects of an inducer on gene expression, adding yet another
means for tuning the behavior of the system. A remarkable feature of our
approach is how accurately our simple model quantitatively describes the mean
response of a wide variety of regulatory contexts. We extend this body of work
by introducing three additional biophysical parameters --
$\Delta\varepsilon_{AI}$, $K_A$, and $K_I$ -- which capture the allosteric
nature of the transcription factor and complement the results shown by Garcia
and Phillips \cite{Garcia2011} and Brewster {\it et al.}\cite{Brewster2014}.

\begin{figure}[h!]
	\centering \includegraphics{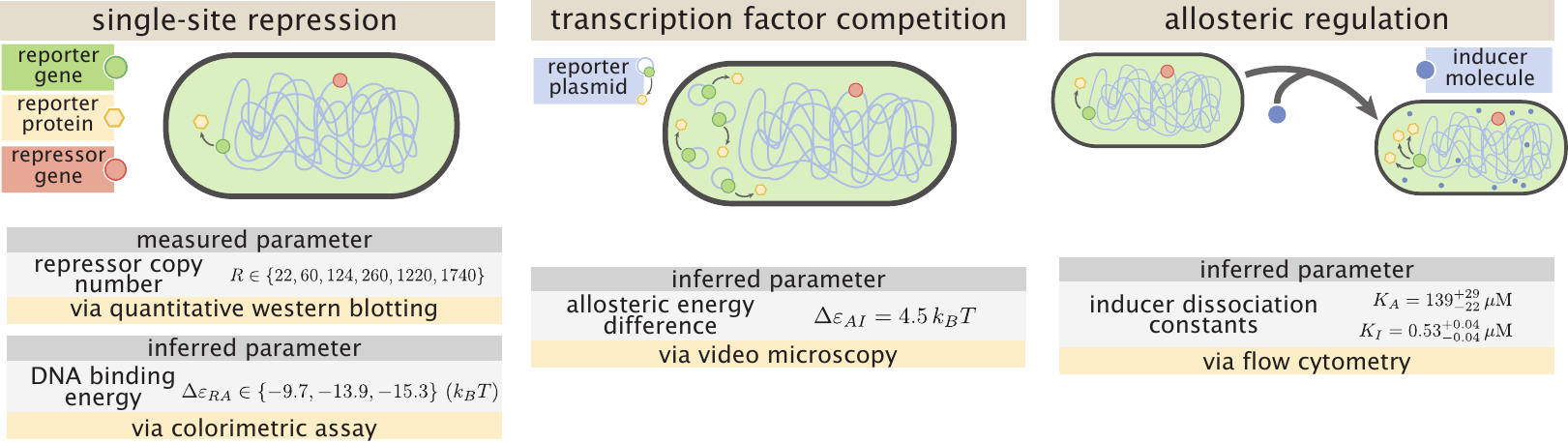}

	\caption{{\bf Understanding the modular components of induction.} Over the past
	decade, we have refined both our experimental control over and theoretical
	understanding of the simple repression architectures. A first round of
	experiments used colorimetric assays and quantitative Western blots to
	investigate how single-site repression is modified by the repressor copy number
	and repressor-DNA binding energy \cite{Garcia2011}. A second round of
	experiments used video microscopy to probe how the copy number of the promoter
	and presence of competing repressor binding sites affect gene expression, and
	we use this data set to determine the free energy difference between the
	repressor's inactive and active conformations \cite{Brewster2014} (see Appendix
	A). Both of the previous experiments characterized the system
	in the absence of an inducer, and in the present work we consider this
	additional important feature of the simple repression architecture. We used
	flow cytometry to determine the inducer-repressor dissociation constants and
	demonstrate that with these parameters we can predict \textit{a priori} the
	behavior of the system for any repressor copy number, DNA binding energy, gene
	copy number, and inducer concentration.} \label{fig_previous_work}
\end{figure}

To test this extension to the theory of transcriptional regulation by simple
repression, we predicted the induction profiles for an array of strains that
could be made using the previously characterized repressor copy number and DNA
binding energies. More specifically, we used modified \textit{lacI} ribosomal
binding sites from Garcia and Phillips \cite{Garcia2011} to generate strains
with mean repressor copy number per cell of $R = 22 \pm 4$, $60 \pm 20$, $124
\pm 30$, $260 \pm 40$, $1220 \pm 160$, and $1740 \pm 340$, where the error
denotes standard deviation of at least three replicates as measured by Garcia
and Phillips \cite{Garcia2011}. We note that repressor copy number $R$ refers to
the number of repressor dimers in the cell, which is twice the number of
repressor tetramers reported by Garcia and Phillips \cite{Garcia2011}; since
both heads of the repressor are assumed to always be either specifically or
non-specifically bound to the genome, the two repressor dimers in each LacI
tetramer can be considered independently. Gene expression was measured using a
Yellow Fluorescent Protein (YFP) gene, driven by a \textit{lacUV5} promoter.
Each of the six repressor copy number variants were paired with the native O1,
O2, or O3 LacI operator \cite{Oehler1994} placed at the YFP transcription start
site, thereby generating eighteen unique strains. The repressor-operator binding
energies (O1 $\Delta\varepsilon_{RA} = -15.3 \pm 0.2~k_BT$, O2
$\Delta\varepsilon_{RA} = -13.9~k_BT \pm 0.2$, and O3 $\Delta\varepsilon_{RA} =
-9.7 \pm 0.1~k_BT$) were previously inferred by measuring the fold-change of the
\textit{lac} system at different repressor copy numbers, where the error arises
from model fitting \cite{Garcia2011}. Additionally, we were able to obtain the
value $\Delta \varepsilon_{AI} = 4.5\ k_BT$ by fitting to previous data as
discussed in Appendix A. We measure fold-change over a range
of known IPTG concentrations $c$, using $n=2$ inducer binding sites per LacI
dimer and approximating the number of non-specific binding sites as the length
in base-pairs of the \textit{E. coli} genome, $N_{NS} = 4.6 \times 10^6$. We
proceed by first inferring the values of the repressor-inducer dissociation
constants $K_A$ and $K_I$ using Bayesian inferential methods as discussed below
\cite{Sivia2006}. When combined with the previously measured parameters within
\eref[eq_fold_change_full], this enables us to predict gene expression for any
concentration of inducer, repressor copy number, and DNA binding energy.

Our experimental pipeline for determining fold-change using flow cytometry is
shown in \fref[fig_experimental_flowchart]. Briefly, cells were grown to
exponential phase, in which gene expression reaches steady state
\cite{Scott2010}, under concentrations of the inducer IPTG ranging between 0 and
$5\,\text{mM}$. We measure YFP fluorescence using flow cytometry and
automatically gate the data to include only single-cell measurements (see
Appendix C). To validate the use of flow cytometry, we
also measured the fold-change of a subset of strains using the established
method of single-cell microscopy (see Appendix D). We
found that the fold-change measurements obtained from microscopy were
indistinguishable from that of flow-cytometry and yielded values for the inducer
binding constants $K_A$ and $K_I$ that were within error.

\begin{figure}[h!]
	\centering \includegraphics{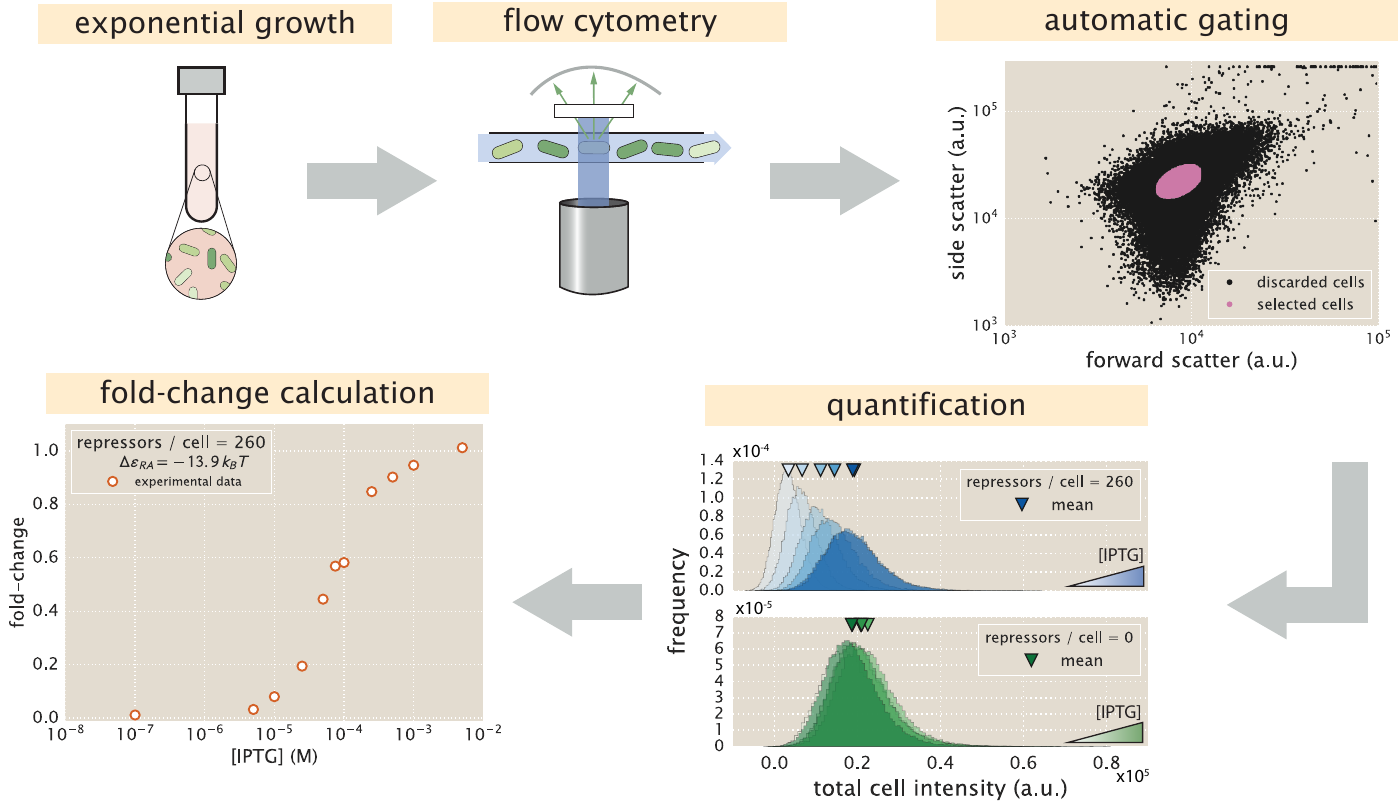}

	\caption{{\bf An experimental pipeline for high-throughput fold-change
		measurements.} Cells are grown to exponential steady state and their
	fluorescence is measured using flow cytometry. Automatic gating methods using
	forward- and side-scattering are used to ensure that all measurements come from
	single cells (see Methods). Mean expression is then quantified at different
	IPTG concentrations (top, blue histograms) and for a strain without repressor
	(bottom, green histograms), which shows no response to IPTG as expected.
	Fold-change is computed by dividing the mean fluorescence in the presence of
	repressor by the mean fluorescence in the absence of repressor.}
\label{fig_experimental_flowchart}
\end{figure}

%%%%%%%%%%%%%%%%%%%%%%%%%%%%%%%%%%%%%%%%%%%%%%%%%%%%%%%%%%%%%%%%%%%%%%%%%%%
\subsection*{Determination of the \textit{in vivo} MWC Parameters}

The three parameters that we tune experimentally are shown in
\fref[fig_O2_R260_fit]\letter{A}, leaving the three allosteric parameters
($\Delta \varepsilon_{AI}$, $K_A$, and $K_I$) to be determined by fitting. Using
previous LacI fold-change data \cite{Brewster2014}, we infer that
$\Delta\varepsilon_{AI} = 4.5~k_BT$ (see Appendix A). Rather
than fitting $K_A$ and $K_I$ to our entire data set of eighteen unique
constructs, we performed a Bayesian parameter estimation on the data from a
single strain with $R=260$ and an O2 operator
($\Delta\varepsilon_{RA}=-13.9~k_BT$ \cite{Garcia2011}) shown in
\fref[fig_O2_R260_fit]\letter{D} (white circles). Using Markov Chain Monte
Carlo, we determine the most likely parameter values to be $K_A=139^{+29}_{-22}
\times 10^{-6} \, \text{M}$ and $K_I=0.53^{+0.04}_{-0.04} \times 10^{-6}\,
\text{M}$, which are the modes of their respective distributions, where the
superscripts and subscripts represent the upper and lower bounds of the
$95^\text{th}$ percentile of the parameter value distributions as depicted in
\fref[fig_O2_R260_fit]\letter{B}. Unfortunately, we are not able to make a
meaningful value-for-value comparison of our parameters to those of earlier
studies \cite{Daber2009, Daber2011a} because of the effects induced by uncertainties in both
gene copy number and transcription factor numbers, the importance of which is illustrated by the plots in
Appendix B. To demonstrate the strength of our
parameter-free model, we then predicted the fold-change for the remaining
seventeen strains with no further fitting (see
\fref[fig_O2_R260_fit]\letter{C}-\letter{E}) together with the specific
phenotypic properties described in
\fref[figInductionCorepressionPhenotypicProperties] and discussed in detail
below (see \fref[fig_O2_R260_fit]\letter{F}-\letter{J}). The shaded regions in
\fref[fig_O2_R260_fit]\letter{C}-\letter{J} denote the 95\% credible regions. An
interesting aspect of our predictions of fold-change is that the width of the
credible regions increases with repressor copy number and inducer concentration
but decreases with the repressor-operator binding strength. Note that the
fold-change \eref[eq_fold_change_full] depends on the product of
$\frac{R}{N_{NS}}e^{-\beta \Delta\varepsilon_{RA}}$ with the MWC parameters
$K_A$, $K_I$, and $\Delta\varepsilon_{AI}$. As a result, strains with small
repressor copy numbers, as well as strains with weak binding energies such as
O3, will necessarily suppress variation in the MWC parameters (see Appendix
E).

We stress that the entire suite of predictions in \fref[fig_O2_R260_fit] is
based upon the induction profile of a single strain. Our ability to make such a
broad range of predictions stems from the fact that our parameters of interest -
such as the repressor copy number and DNA binding energy - appear as distinct physical
parameters within our model. While the single data set in
\fref[fig_O2_R260_fit]\letter{D} could also be fit using a Hill
function, such an analysis would be unable to predict any of the other curves in
the figure. Phenomenological expressions such as the Hill function
can describe data, but lack predictive power and are
thus unable to build our intuition, design \textit{de novo} input-output
functions, or guide future experiments.

\begin{figure}[H]
	\centering \includegraphics{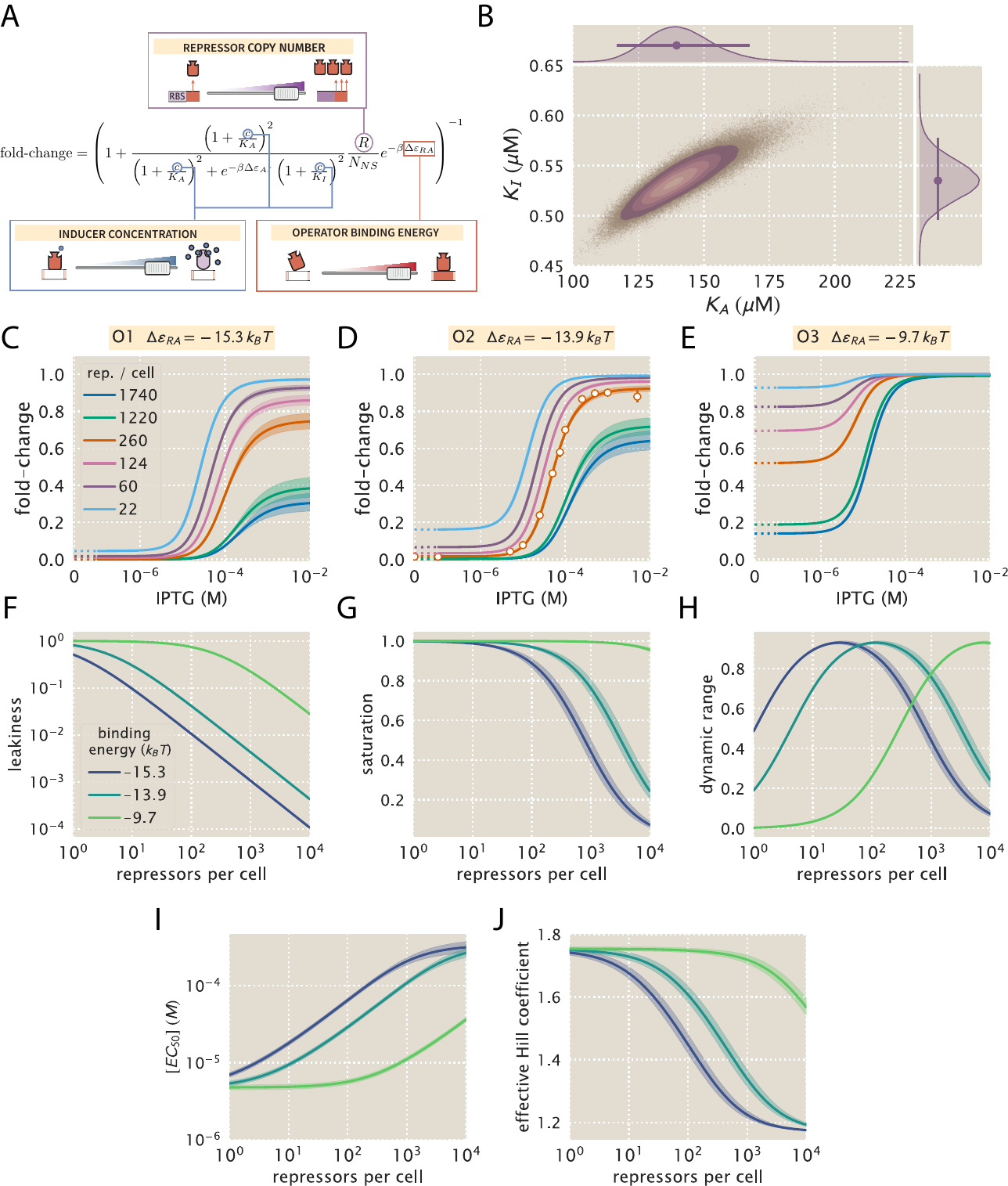}

	\caption{\textbf{Predicting induction profiles for different biological control
		parameters.} \letterParen{A} We can quantitatively tune $R$ via ribosomal
	binding site (RBS) modifications, $\Delta\varepsilon_{RA}$ by mutating the
	operator sequence, and $c$ by adding different amounts of IPTG to the growth
	medium. \letterParen{B} Previous experiments have characterized the $R$,
	$N_{NS}$, $\Delta\varepsilon_{RA}$, and $\Delta\varepsilon_{AI}$ parameters
	(see \fref[fig_previous_work]), leaving only the unknown dissociation constants
	$K_A$ and $K_I$ between the inducer and the repressor in the active and
	inactive states, respectively. These two parameters can be inferred using
	Bayesian parameter estimation from a single induction curve. \letterParen{C-E}
	Predicted IPTG titration curves for different repressor copy numbers and
	operator strengths. Titration data for the O2 strain (white circles in Panel
	\letter{D}) with $R=260$, $\Delta\varepsilon_{RA} = -13.9~k_BT$, $n=2$, and
	$\Delta\varepsilon_{AI}=4.5~k_BT$ can be used to determine the thermodynamic
	parameters $K_A=139^{+29}_{-22} \times 10^{-6} \, \text{M}$ and
	$K_I=0.53^{+0.04}_{-0.04} \times 10^{-6}\, \text{M}$ (orange line). The
	remaining solid lines predict the fold-change \eref[eq_fold_change_full] for
	all other combinations of repressor copy numbers (shown in the legend) and
	repressor-DNA binding energies corresponding to the O1 operator ($-15.3~k_B
	T$), O2 operator ($-13.9~k_B T$), and O3 operator ($-9.7~k_B T$). Error bars
	of experimental data show the standard error of the mean (eight or more
	replicates) when this error is not smaller than the diameter of the data point.
	The shaded regions denote the 95\% credible region, although the credible
	region is obscured when it is thinner than the curve itself. To display the
	measured fold-change in the absence of inducer, we alter the scaling of the
	$x$-axis between $0$ and $10^{-7}$ M to linear rather than logarithmic, as
	indicated by a dashed line. Additionally, our model allows us to investigate
	key phenotypic properties of the induction profiles (see
	\fref[figInductionCorepressionPhenotypicProperties]\letter{B}). Specifically,
	we show predictions for the \letterParen{F} leakiness, \letterParen{G}
	saturation, \letterParen{H} dynamic range, \letterParen{I} $[EC_{50}]$, and
	\letterParen{J} effective Hill coefficient of the induction profiles.}
\label{fig_O2_R260_fit}
\end{figure}

%%%%%%%%%%%%%%%%%%%%%%%%%%%%%%%%%%%%%%%%%%%%%%%%%%%%%%%%%%%%%%%%%%%%%%%%%%%
\subsection*{Comparison of Experimental Measurements with Theoretical Predictions}

We tested the predictions shown in \fref[fig_O2_R260_fit] by measuring the
fold-change induction profiles using strains that span this broad range in
repressor copy numbers and repressor binding energies as characterized in
\cite{Garcia2011}, and inducer concentrations spanning several orders of
magnitude. The results, shown in \fref[fig_O2_R260_pred_data], demonstrate very
good agreement between theory and experiment across all of our strains. We note,
however, that there was an apparently systematic shift in the O3 $\Delta\varepsilon_{RA} = -9.7\ k_BT$ strains
(\fref[fig_O2_R260_pred_data]C) and all of the $R=1220$ and $R =1740$ strains.
This may be partially due to imprecise previous determinations of their
$\Delta\varepsilon_{RA}$ and $R$ values. By performing a global fit where we
infer all parameters including the repressor copy number $R$ and the binding
energy $\Delta\varepsilon_{RA}$, we found better agreement for these particular
strains, although a discrepancy in the steepness of the response for all O3
strains remains (see Appendix F). As an additional test
of our model, we also considered strains using the synthetic Oid operator which
exhibits stronger repression, $\Delta\varepsilon_{RA}=-17~k_B T$
\cite{Garcia2011}, than the O1, O2, and O3 operators. We found that we were
unable to measure the strongly repressed strains accurately by flow cytometry.
However, for the data we collected, we found that the MWC description was consistent to
within $1~k_B T$ of the binding energy previously reported (see Appendix
G for more details).

To ensure that the agreement between our predictions and data is not an accident
of the strain we chose to perform our fitting, we explored the effects of using
each of our other strains to estimate $K_A$ and $K_I$. As shown in Appendix
H and \fref[fig_O2_R260_pred_data]\letter{D}, the
inferred values of $K_A$ and $K_I$ depend very minimally upon which strain is
chosen, demonstrating that these parameter values are highly robust. As
previously mentioned, we performed a global fit using the data from all eighteen
strains for the following parameters: the inducer dissociation constants $K_A$
and $K_I$, the repressor copy numbers $R$, and the repressor DNA binding energy
$\Delta\varepsilon_{RA}$ (see Appendix F). This global
fit led to very similar parameter values, lending strong support for our
quantitative understanding of induction in the simple repression architecture.
For the remainder of the text we proceed using our analysis on the strain with
$R=260$ repressors and an O2 operator.

\begin{figure}[H]
	\centering \includegraphics{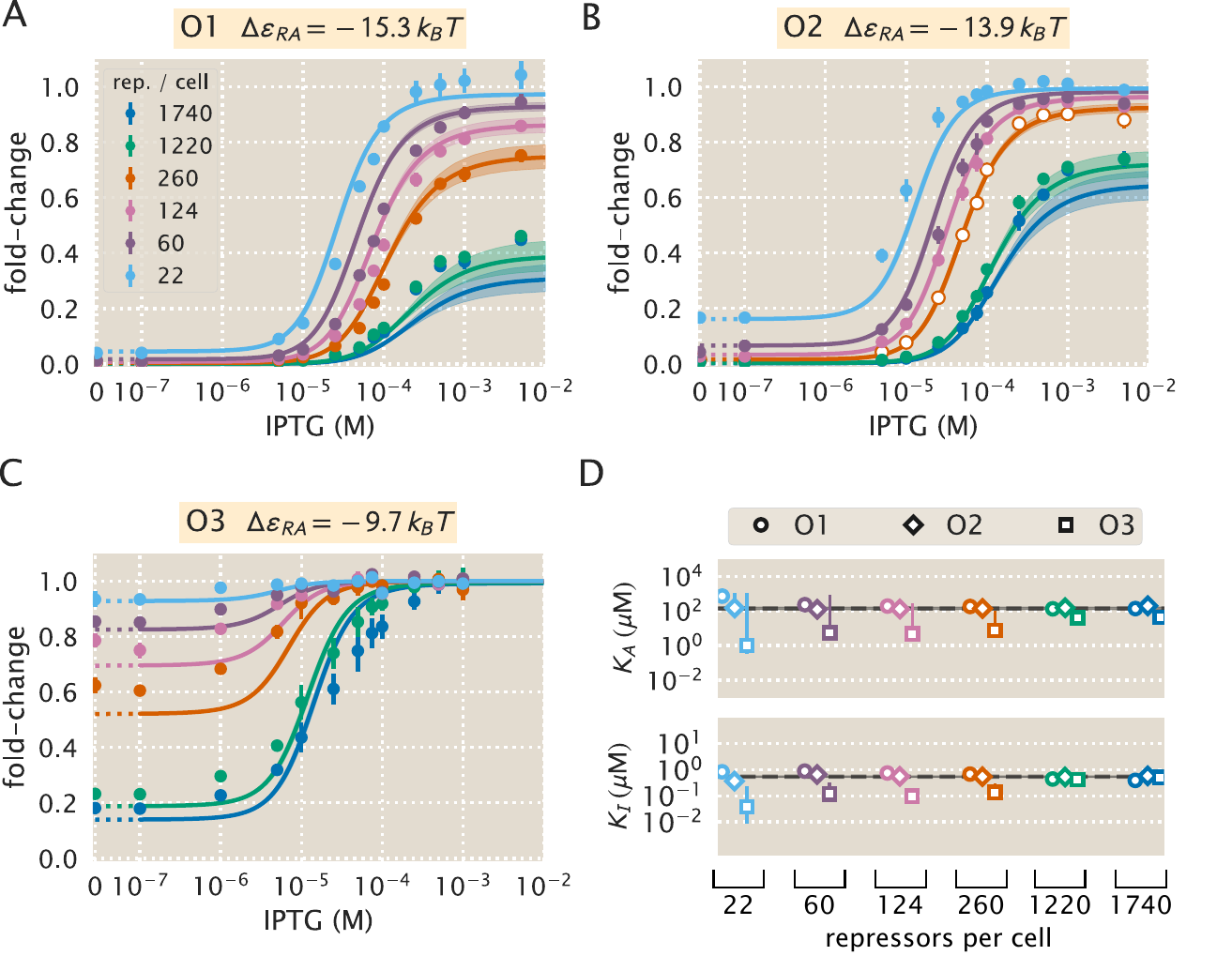}

	\caption{\textbf{Comparison of predictions against measured and inferred
	data.} Flow cytometry measurements of fold-change over a range of IPTG
	concentrations for \letterParen{A} O1, \letterParen{B} O2, and \letterParen{C}
	O3 strains at varying repressor copy numbers, overlaid on the predicted
	responses. Error bars of the experimental data show the standard error of the
	mean (eight or more replicates). As discussed in \fref[fig_O2_R260_fit], all
	of the predicted induction curves were created prior to measurement by
	inferring the MWC parameters using a single data set (O2 $R=260$, shown by
	white circles in Panel \letter{B}). The predictions may therefore depend upon
	which strain is used to infer the parameters. \letterParen{D} The inferred
	parameter values of the dissociation constants $K_A$ and $K_I$ using any of
	the eighteen strains instead of the O2 $R=260$ strain. Nearly identical
	parameter values are inferred from each strain, demonstrating that the same
	set of induction profiles would have been predicted regardless of which strain
	was chosen. The points show the mode and the error bars denote the $95\%$
	credible region of the parameter value distribution. Error bars not visible
	are smaller than the size of the marker.} 	\label{fig_O2_R260_pred_data}
\end{figure}

%%%%%%%%%%%%%%%%%%%%%%%%%%%%%%%%%%%%%%%%%%%%%%%%%%%%%%%%%%%%%%%%%%%%%%%%%%%
\subsection*{Predicting the Phenotypic Traits of the Induction Response}

Rather than measuring the full induction response of a system, a subset of the properties shown in \fref[figInductionCorepressionPhenotypicProperties], namely, the leakiness, saturation, dynamic range, $[EC_{50}]$, and effective Hill coefficient, may be of greater interest. For example, synthetic
biology is often focused on generating large responses (i.e. a large dynamic
range) or finding a strong binding partner (i.e. a small $[EC_{50}]$)
\cite{Brophy2014, Shis2014}. While these properties are all individually informative, when taken together they capture the essential features of
the induction response. We reiterate that a Hill function approach cannot predict
these features \textit{a priori} and furthermore requires fitting each curve
individually. The MWC model, on the other hand, enables us to quantify how each
trait depends upon a single set of physical parameters as shown by
\fref[fig_O2_R260_fit]\letter{F}-\letter{J}.

We define these five phenotypic traits using expressions derived from the model, \eref[eq_fold_change_full]. These results build upon extensive work by Martins and Swain, who computed many such properties for ligand-receptor binding within the MWC model \cite{Martins2011}. We begin by analyzing the
leakiness, which is the minimum fold-change observed in the absence of
ligand, given by
\begin{linenomath}
\begin{align} \label{eqLeakiness}
\text{leakiness} &= \foldchange(c=0) \nonumber\\
&= \left(
	1+\frac{1}{1+e^{-\beta \Delta \varepsilon_{AI} }}\frac{R}{N_{NS}}e^{-\beta \Delta\varepsilon_{RA}} \right)^{-1},
\end{align}
\end{linenomath}
and the saturation, which is the maximum fold change observed in the presence of saturating ligand,
\begin{linenomath}
\begin{align} \label{eqSaturation}
\text{saturation} &= \foldchange(c \to \infty) \nonumber\\
&= \left(
	1+\frac{1}{1+e^{-\beta \Delta \varepsilon_{AI} } \left(\frac{K_A}{K_I}\right)^n }\frac{R}{N_{NS}}e^{-\beta \Delta\varepsilon_{RA}} \right)^{-1}.
\end{align}
\end{linenomath}

Systems that minimize leakiness repress strongly in the absence of effector while
systems that maximize saturation have high expression in the presence of
effector. Together, these two properties determine the dynamic range of a
system's response, which is given by the difference
\begin{equation} \label{eqDynamicRangeDef}
	\text{dynamic range} = \text{saturation} - \text{leakiness}.
\end{equation}
These three properties are shown in \fref[fig_O2_R260_fit]\letter{F}-\letter{H}.
We discuss these properties in greater detail in Appendix
I. For example, we compute the number of repressors $R$
necessary to evoke the maximum dynamic range and demonstrate that the magnitude
of this maximum is independent of the repressor-operator binding energy
$\Delta\varepsilon_{RA}$. \fref[fig_properties_data]\letter{A}-\letter{C} show
that the measurements of these three properties, derived from the fold-change
data in the absence of IPTG and the presence of saturating IPTG, closely match
the predictions for all three operators.

\begin{figure}[h!]
	\centering \includegraphics{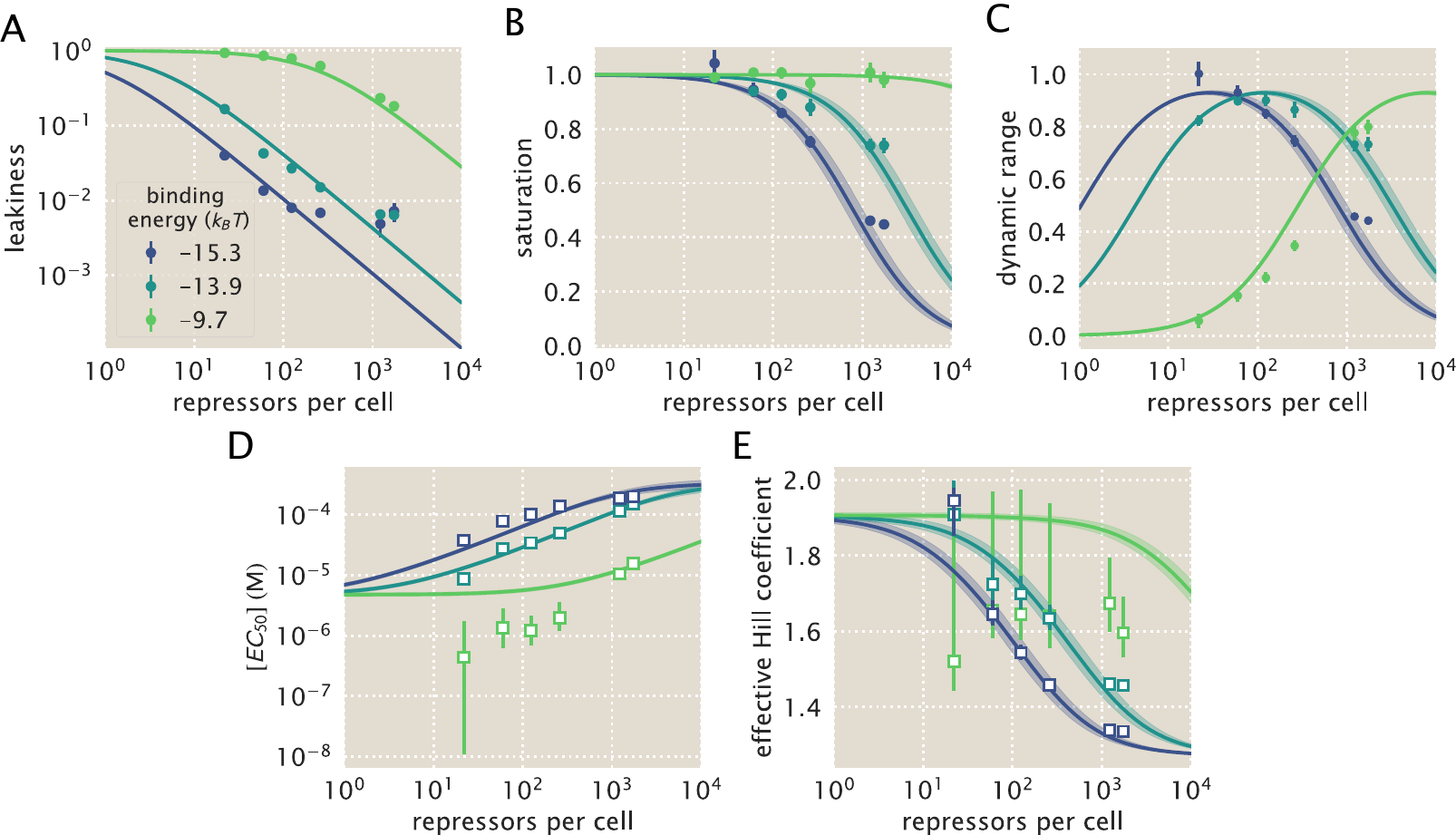}

	\caption{\textbf{Predictions and experimental measurements of key properties of
		induction profiles.} Data for the \letterParen{A} leakiness, \letterParen{B}
	saturation, and \letterParen{C} dynamic range are obtained from fold-change
	measurements in \fref[fig_O2_R260_pred_data] in the absence of IPTG and at
	saturating concentrations of IPTG. The three repressor-operator binding
	energies in the legend correspond to the O1 operator ($-15.3~k_B T$), O2
	operator ($-13.9~k_B T$), and O3 operator ($-9.7~k_B T$). Both the
	\letterParen{D} $[EC_{50}]$ and \letterParen{E} effective Hill coefficient are
	inferred by individually fitting each operator-repressor pairing in
	\fref[fig_O2_R260_pred_data]\letter{A}-\letter{C} separately to
	\eref[eq_fold_change_full] in order to smoothly interpolate between the data
	points. Error bars for \letter{A}-\letter{C} represent the standard error of
	the mean for eight or more replicates; error bars for \letter{D}-\letter{E}
	represent the 95\% credible region for the parameter found by propagating the
	credible region of our estimates of $K_A$ and $K_I$ into
	\eref[ec50][effectiveHill].} \label{fig_properties_data}
\end{figure}

Two additional properties of induction profiles are the $[EC_{50}]$ and
effective Hill coefficient, which determine the range of inducer concentration
in which the system's output goes from its minimum to maximum value. The
$[EC_{50}]$ denotes the inducer concentration required to generate a system
response \eref[eq_fold_change_full] halfway between its minimum and maximum
value,
\begin{equation} \label{ec50}
\foldchange(c = [EC_{50}]) = \frac{\text{leakiness} + \text{saturation}}{2}.
\end{equation}
The effective Hill coefficient $h$, which quantifies the steepness of the curve at the $[EC_{50}]$ \cite{Marzen2013}, is given by
\begin{equation} \label{effectiveHill}
h = \left( 2 \frac{d}{d \log c} \left[ \log \left( \frac{ \foldchange(c) - \text{leakiness}}{\text{dynamic range}} \right) \right] \right)_{c = [EC_{50}]}.
\end{equation}
\fref[fig_O2_R260_fit]\letter{I}-\letter{J} shows how the $[EC_{50}]$ and
effective Hill coefficient depend on the repressor copy number. In Appendix
I, we discuss the analytic forms of these two properties
as well as their dependence on the repressor-DNA binding energy.

\fref[fig_properties_data]\letter{D}-\letter{E} show the estimated values of the
$[EC_{50}]$ and the effective Hill coefficient overlaid on the theoretical
predictions. Both properties were obtained by fitting \eref[eq_fold_change_full]
to each individual titration curve and computing the $[EC_{50}]$ and effective
Hill coefficient using \eref[ec50] and \eref[effectiveHill], respectively. We
find that the predictions made with the single strain fit closely match those
made for each of the strains with O1 and O2 operators, but the predictions for
the O3 operator are markedly off. The large, asymmetric error bars for the O3
$R=22$ strain arise from its nearly flat response, where the lack of dynamic
range makes it impossible to determine the value of the inducer dissociation
constants $K_A$ and $K_I$; consequently the determination of $[EC_{50}]$ is
accompanied with significant uncertainty.

%%%%%%%%%%%%%%%%%%%%%%%%%%%%%%%%%%%%%%%%%%%%%%%%%%%%%%%%%%%%%%%%%%%%%%%%%%%
\subsection*{Data Collapse of Induction Profiles}

Our primary interest heretofore was to determine the system response at a
specific inducer concentration, repressor copy number, and repressor-DNA binding
energy. We now flip this question on its head and ask: given a specific value of
the fold-change, what combination of parameters will give rise to this desired
response? In other words, what trade-offs between the parameters of the system
will give rise to the same mean cellular output? These are key questions both
for understanding how the system is governed and for engineering specific
responses in a synthetic biology context. To this end, we rewrite
\eref[eq_fold_change_full] as a Fermi function,
\begin{equation}\label{eq_free_energy_definition}
\foldchange= \frac{1}{1+e^{-\beta F(c)}},
\end{equation}
where $F(c)$ is the free energy of the repressor binding to the operator of
interest relative to the unbound operator state \cite{Keymer2006, Swem2008,
	Phillips2015a}, which is given by
\begin{equation}\label{eq_free_energy_MWC_parameters}
F(c) = - k_BT \left( \log \frac{\left(1+\frac{c}{K_A}\right)^n}{\left(1+\frac{c}{K_A}\right)^n+e^{-\beta \Delta\varepsilon_{AI} }\left(1+\frac{c}{K_I}\right)^n} + \log \frac{R}{N_{\text{NS}}} - \frac{\Delta\varepsilon_{RA}}{k_BT} \right).
\end{equation}
The first term in the parenthesis denotes the contribution from the inducer
concentration, the second the effect of the repressor copy number, and the last
the repressor-operator binding energy. We note that elsewhere, this free energy
has been dubbed the Bohr parameter since such families of curves are analogous
to the shifts in hemoglobin binding curves at different pHs known as the Bohr
effect \cite{Mirny2010, Phillips2015a, Einav2016}.

Instead of analyzing each induction curve individually, the free energy provides
a natural means to simultaneously characterize the diversity in our eighteen
induction profiles. \fref[fig_datacollapse]\letter{A} demonstrates how the
various induction curves from \fref[fig_O2_R260_fit]\letter{C}-\letter{E} all
collapse onto a single master curve, where points from every induction profile
that yield the same fold-change are mapped onto the same free energy.
\fref[fig_datacollapse]\letter{B} shows this data collapse for the 216 data
points in \fref[fig_O2_R260_pred_data]\letter{A}-\letter{C}, demonstrating the
close match between the theoretical predictions and experimental measurements
across all eighteen strains.

There are many different combinations of parameter values that can result in the
same free energy as defined in \eref[eq_free_energy_MWC_parameters]. For
example, suppose a system originally has a fold-change of 0.2 at a specific
inducer concentration, and then operator mutations increase the
$\Delta\varepsilon_{RA}$ binding energy. While this serves to initially increase
both the free energy and the fold-change, a subsequent increase in the
repressor copy number could bring the cell back to the original fold-change
level. Such trade-offs hint that there need not be a single set of parameters
that evoke a specific cellular response, but rather that the cell explores a
large but degenerate space of parameters with multiple, equally valid paths.

\begin{figure}[ht]
	\centering \includegraphics{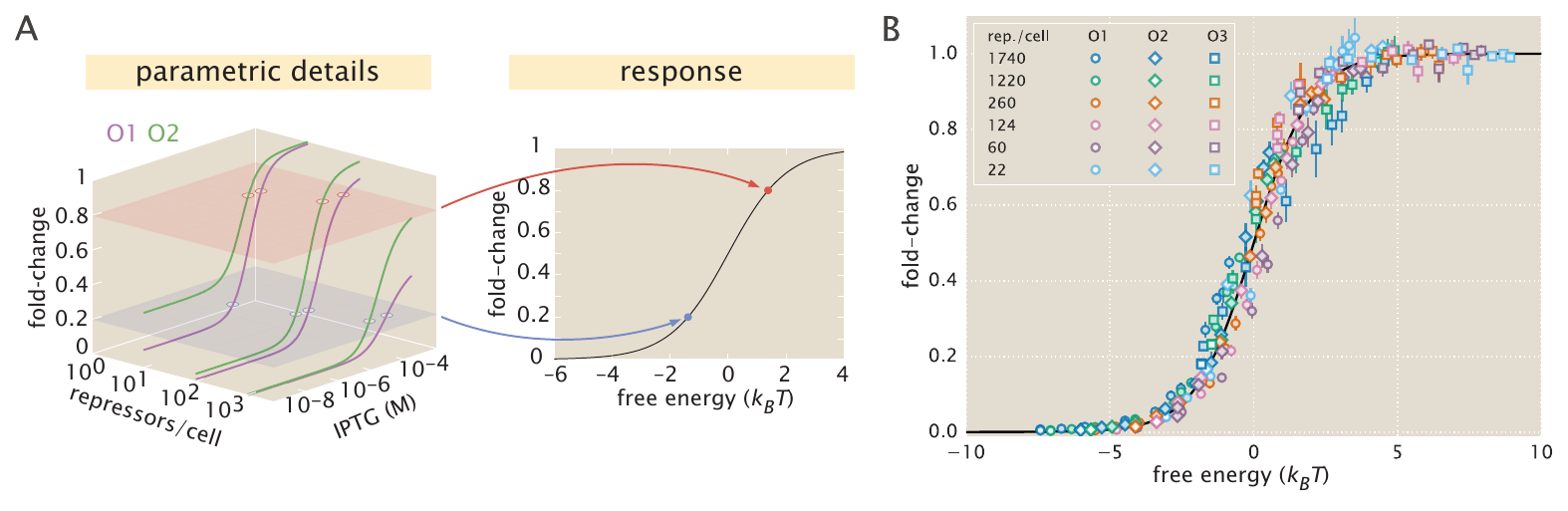}

	\caption{\textbf{Fold-change data from a broad collection of different strains
		collapse onto a single master curve.} \letterParen{A} Any combination of
	parameters can be mapped to a single physiological response (i.e. fold-change)
	via the free energy, which encompasses the parametric details of the model.
	\letterParen{B} Experimental data from \fref[fig_O2_R260_pred_data] collapse
	onto a single master curve as a function of the free energy
	\eref[eq_free_energy_MWC_parameters]. The free energy for each strain was
	calculated from \eref[eq_free_energy_MWC_parameters] using $n=2$,
	$\Delta\varepsilon_{AI}=4.5~k_BT$, $K_A=139 \times 10^{-6} \, \text{M}$,
	$K_I=0.53 \times 10^{-6}\, \text{M}$, and the strain specific $R$ and
	$\Delta\varepsilon_{RA}$. All data points represent the mean and error bars are
	the standard error of the mean for eight or more replicates.}
\label{fig_datacollapse}
\end{figure}

	%%%%%%%%%%%%%%%%%%%%%%%%%%%%%%%%%%%%%%%%%%%%%%%%%%%%%%%%%%%%%%%%%%%%%%%%%%%
\section*{Discussion}

Since the early work by Monod, Wyman, and Changeux \cite{Monod1963, MONOD1965},
a broad list of different biological phenomena have been tied to the existence
of macromolecules that switch between inactive and active states. Examples can
be found in a wide variety of cellular processes that include ligand-gated ion
channels \cite{Auerbach2012}, enzymatic reactions \cite{Einav2016, Velyvis2007},
chemotaxis \cite{Keymer2006}, quorum sensing \cite{Swem2008}, G-protein coupled receptors \cite{Canals2012},
physiologically important proteins \cite{Levantino2012a, Milo2007}, and beyond.
One of the most ubiquitous examples of allostery is in the context of gene
expression, where an array of molecular players bind to transcription factors to
either aid or deter their ability to regulate gene activity
\cite{Huang2011,Li2014}. Nevertheless, no definitive study has been made of the
applicability of the MWC model to transcription factor function, despite the
clear presence of different conformational states in their structures in the
presence and absence of signaling molecules \cite{Lewis1996}. A central goal of
this work was to assess whether a thermodynamic MWC model can provide an
accurate input-output function for gene regulation by allosteric transcription
factors.

Others have developed quantitative models describing different aspects of
allosteric regulatory systems. Martins and Swain analytically derived
fundamental properties of the MWC model, including the leakiness and dynamic
range described in this work, noting the inherent trade-offs in these properties
when tuning the microscopic parameters of the model \cite{Martins2011, Marzen2013}. Work in
the Church and Voigt labs, among others, has expanded on the availability of
allosteric circuits for synthetic biology
\cite{Rogers2015,Moon2012,Lutz1997,Rohlhill2017}. Recently, Daber \textit{et
	al.} theoretically explored the induction of simple repression within the MWC
model \cite{Daber2009} and experimentally measured how mutations alter the
induction profiles of transcription factors \cite{Daber2011a}. Vilar and Saiz
considered the broad range of interactions in inducible \textit{lac}-based
systems including the effects of oligomerization and DNA folding on
transcription factor induction \cite{Leonor2008,Vilar2013}. Other work has
attempted to use the \textit{lac} system to reconcile \textit{in vitro} and
\textit{in vivo} measurements \cite{Tungtur2011, Sochor2014}. Although this body
of work has done much to improve our understanding of allosteric transcription
factors, there has remained a disconnect between model and experiment. In order
to rigorously test a model's applicability to natural systems, the model's
predictions must be weighed against data from precise experiments specifically
designed to test those predictions.

Here, we expand upon this body of work by generating a predictive model of
allosteric transcriptional regulation and then testing the model against a
thorough set of experiments using well-characterized regulatory components.
Specifically, we used the MWC model to build upon and refine a well-established
thermodynamic model of transcriptional regulation\cite{Bintu2005, Garcia2011},
allowing us to compose the model from a minimal set of biologically meaningful
parameters. This minimal model captures the key players of transcriptional
regulation -- namely the repressor copy number, the DNA binding energy, and the
concentration of inducer -- and enables us to predict how the system will behave
when we change each of these parameters. We tested these predictions on a
range of strains whose repressor copy number spanned two orders of magnitude and
whose DNA binding affinity spanned 6 $k_BT$. We argue that one would not be able
to generate such a wide array of predictions by using a Hill function, which
abstracts away the biophysical meaning of the parameters into phenomenological
parameters \cite{Forsen1995}.

Specifically, we tested our model in the context of a \textit{lac}-based simple
repression system by first determining the allosteric dissociation constants
$K_A$ and $K_I$ from a single induction data set (O2 operator with binding
energy $\Delta \varepsilon_{RA} = -13.9~k_BT$ and repressor copy number $R =
260$) and then using these values to make parameter-free predictions of the
induction profiles for seventeen other strains where $\Delta \varepsilon_{RA}$
and $R$ were varied significantly (see \fref[fig_O2_R260_fit]). We next measured
the induction profiles of these seventeen strains using flow cytometry and found
that our predictions consistently and accurately captured the primary features
for each induction data set, as shown in
\fref[fig_O2_R260_pred_data]\letter{A}-\letter{C}. Surprisingly, we find that
the inferences for the repressor-inducer dissociation constants that would have
been derived from any other single strain (instead of the O2 operator with
$R=260$) would have resulted in nearly identical predictions (see
\fref[fig_O2_R260_pred_data]\letter{D} and Appendix
H). This suggests that a few carefully chosen
measurements can lead to a deep quantitative understanding of how simple
regulatory systems work without requiring an extensive sampling of strains that
span the parameter space. Moreover, the fact that we could consistently achieve
reliable predictions after fitting only two free parameters stands in contrast to
the common practice of fitting several free parameters simultaneously, which can
nearly guarantee an acceptable fit provided that the model roughly resembles the
system response, regardless of whether the details of the model are tied
to any underlying molecular mechanism.

Beyond observing changes in fold-change as a function of effector concentration,
our application of the MWC model allows us to explicitly predict the values of
the induction curves' key parameters, namely, the leakiness, saturation, dynamic
range, $[EC_{50}]$, and the effective Hill coefficient (see
\fref[fig_properties_data]). This allows us to quantify the unique traits of
each set of strains examined here. Strains using the O1 operator consistently
have a low leakiness value, a consequence of its strong binding energy. The
saturation values for these strains, however, vary significantly with $R$. This
trend is reversed for strains using O3, which has the weakest binding energy of
our constructs. Leakiness values for constructs using O3 vary strongly with $R$,
but their saturation values approach 1 regardless of $R$. Strains with the
intermediate O2 binding energy have both a leakiness and saturation that vary
markedly with $R$. For both the O1 and O2 data sets, our model also accurately
predicts the effective Hill coefficient and $[EC_{50}]$, though these
predictions for O3 are noticeably less accurate. While performing a global fit
for all model parameters marginally improves the prediction for O3 (see Appendix
F), we are still unable to accurately predict the
effective Hill coefficient or the $[EC_{50}]$, though the uncertainties in these
two parameters are really an inheritance from the consistent difference between
the theoretical and measured sharpness of the induction response seen in
\fref[fig_O2_R260_pred_data]\letter{C}.

Because this model allows us to derive expressions for individual features of
induction curves, we are able to examine how these features may be tuned by
careful selection of system parameters. \fref[fig_properties_data] shows how
each of the induction curves' key features vary as a function of $\Delta
\varepsilon_{RA}$ and $R$, which makes it possible to select desired properties
from among the possible phenotypes available to the system. For instance, it is
possible to obtain a high dynamic range using fewer than 100 repressors if the
binding energy is strong. As an example of the constraints inherent to the
system, one cannot design a strain with a leakiness of 0.1 and a saturation of
0.4 by only varying the repressor copy number and repressor-operator binding
affinity, since these two properties are coupled by
\eref[eqLeakiness][eqSaturation]. Achieving this particular behavior would
require changing the ratio $K_A/K_I$ of repressor-inducer dissociation
constants, as may be done by mutating the repressor's inducer binding pocket.

The dynamic range, which is of considerable interest when designing or
characterizing a genetic circuit, is revealed to have an interesting property:
although changing the value of $\Delta \varepsilon_{RA}$ causes the dynamic
range curves to shift to the right or left, each curve has the same shape and in
particular the same maximum value. This means that strains with strong or weak
binding energies can attain the same dynamic range when the value of $R$ is
tuned to compensate for this energy. This feature is not immediately apparent
from the IPTG induction curves, which show very low dynamic ranges for several
of the O1 and O3 strains. Without the benefit of models that can predict such
phenotypic traits, efforts to engineer genetic circuits with allosteric
transcription factors must rely on trial and error to achieve specific responses
\cite{Rogers2015,Rohlhill2017}. This is a compelling example showing that our
predictive modeling approach has a significant advantage over descriptive
models.

To our knowledge this is the first work of its kind in which a single family of
parameters is demonstrated to predict a vast range of induction curves with
qualitatively different behaviors. One of the demanding criteria of our approach
is that a small set of parameters must consistently describe data from a diverse
collection of data sets taken using distinct methods such as Miller assays and
bulk and single-cell fluorescence experiments to measure fold-change (see
Appendices C and G), as well as
quantitative Western blots \cite{Garcia2011} and binomial partitioning methods
to count repressors \cite{Brewster2014,Rosenfeld2005}. Furthermore, we build off
of our previous studies that use the simple repression architecture and we
demand that the parameters derived from these studies account for constructs
that are integrated into the chromosome, plasmid-borne, and even for cases where
there are competing binding sites to take repressors out of circulation
\cite{Garcia2011, Brewster2014} (see Appendix B) or where
there are multiple operators to allow DNA looping \cite{Boedicker2013a}. The
resulting model not only predicts the individual titration profiles as a
function of IPTG, but describes key properties of the response. The general
agreement with the entire body of work presented here demonstrates that our
model captures the underlying mechanism governing simple repression. We are
unaware of any comparable study in transcriptional regulation that demands one
predictive framework cover such a broad array of regulatory situations.

Despite the diversity observed in the induction profiles of each of our strains,
our data are unified by their reliance on fundamental biophysical parameters. In
particular, we have shown that our model for fold-change can be rewritten in
terms of the free energy \eref[eq_free_energy_MWC_parameters], which encompasses
all of the physical parameters of the system. This has proven to be an
illuminating technique in a number of studies of allosteric proteins
\cite{Sourjik2002,Keymer2006,Swem2008}. Although it is experimentally
straightforward to observe system responses to changes in effector concentration
$c$, framing the input-output function in terms of $c$ can give the misleading
impression that changes in system parameters lead to fundamentally altered
system responses. Alternatively, if one can find the ``natural variable" that
enables the output to collapse onto a single curve, it becomes clear that the
system's output is not governed by individual system parameters, but rather the
contributions of multiple parameters that define the natural variable.

When our fold-change data are plotted against the respective free energies for
each construct, they collapse cleanly onto a single curve (see
\fref[fig_datacollapse]). This enables us to analyze how parameters can
compensate each other. For example, we may wish to determine which combinations
of parameters result in a system that is strongly repressed (free energy $F(c)
\leq -5~k_B T$). We know from our understanding of the induction phenomenon that
strong repression is most likely to occur at low values of $c$. However, from
\eref[eq_free_energy_MWC_parameters] we can clearly see that increases in the
value of $c$ can be compensated by an increase in the number of repressors $R$,
a decrease in the binding energy $\Delta \varepsilon_{RA}$ (i.e. stronger
binding), or some combination of both. Likewise, while the system tends to
express strongly ($F(c) \geq 5~k_B T$) when $c$ is high, one could design a
system that expresses strongly at low values of $c$ by reducing $R$ or
increasing the value of $\Delta \varepsilon_{RA}$. As a concrete example, given
a concentration $c = 10^{-5}\,\text{M}$, a system using the O1 operator ($\Delta
\varepsilon_{RA} = -15.3~k_B T$) requires 745 or more repressors for $F(c) \leq
-5~k_B T$, while a system using the weaker O3 operator ($\Delta \varepsilon_{RA}
= -9.7~k_B T$) requires $2 \times 10^5$ or more repressors for $F(c) \leq
-5~k_BT$.

While our experiments validated the theoretical predictions in the case of
simple repression, we expect the framework presented here to apply much more
generally to different biological instances of allosteric regulation. For
example, we can use this model to explore different regulatory configurations
such as corepression, activation, and coactivation, each of which are found in
\textit{E. coli} (see Appendix J). This work can also
serve as a springboard to characterize not just the mean but the full gene
expression distribution and thus quantify the impact of noise on this system
\cite{eldar2010}. Another extension of this approach would be to theoretically
predict and experimentally verify whether the repressor-inducer dissociation
constants $K_A$ and $K_I$ or the energy difference $\Delta \varepsilon_{AI}$
between the allosteric states can be tuned by making single amino acid
substitutions in the transcription factor \cite{Daber2011a, Phillips2015a}.
Finally, we expect that the kind of rigorous quantitative description of the
allosteric phenomenon provided here will make it possible to construct
biophysical models of fitness for allosteric proteins similar to those already
invoked to explore the fitness effects of transcription factor binding site
strengths and protein stability \cite{Gerland2002, Berg2004, Zeldovich2008}.

To conclude, we find that our application of the MWC model provides an accurate,
predictive framework for understanding simple repression by allosteric
transcription factors. To reach this conclusion, we analyzed the model in the
context of a well-characterized system, in which each parameter had a clear
biophysical meaning. As many of these parameters had been measured or inferred
in previous studies, this gave us a minimal model with only two free parameters
which we inferred from a single data set. We then accurately predicted the
behavior of seventeen other data sets in which repressor copy number and
repressor-DNA binding energy were systematically varied. In addition, our model
allowed us to understand how key properties such as the leakiness, saturation,
dynamic range, $[EC_{50}]$, and effective Hill coefficient depended upon the
small set of parameters governing this system. 
%We further note that our model could be extended to encompass more complex
%systems involving multiple promoters or competitor binding sites, as well as
%different architectures such as corepression and activation.
Finally, we show that by framing inducible simple
repression in terms of free energy, the data from all of our experimental
strains collapse cleanly onto a single curve, illustrating the many ways in
which a particular output can be targeted. In total, these results show that a
thermodynamic formulation of the MWC model supersedes phenomenological
fitting functions for
understanding transcriptional regulation by allosteric proteins.

	%%%%%%%%%%%%%%%%%%%%%%%%%%%%%%%%%%%%%%%%%%%%%%%%%%%%%%%%%%%%%%%%%%%%%%%%%%%
\section*{Methods} \label{section_methods}

\subsection*{Bacterial Strains and DNA Constructs}

All strains used in these experiments were derived from \textit{E. coli} K12
MG1655 with the \textit{lac} operon removed, adapted from those created and
described in \cite{Garcia2011, Garcia2011B}. Briefly, the operator variants and
YFP reporter gene were cloned into a pZS25 background which contains a
\textit{lacUV5} promoter that drives expression as is shown in
\fref[fig_polymerase_repressor_states]. These constructs carried a kanamycin
resistance gene and were integrated into the \textit{galK} locus of the
chromosome using $\lambda$ Red recombineering \cite{Sharan2009}. The
\textit{lacI} gene was constitutively expressed via a
P$_\mathrm{LtetO\hbox{-}1}$ promoter \cite{Lutz1997}, with ribosomal binding
site mutations made to vary the LacI copy number as described in
\cite{Salis2009} using site-directed mutagenesis (Quickchange II; Stratagene),
with further details in \cite{Garcia2011}. These {\it lacI} constructs carried a
chloramphenicol resistance gene and were integrated into the \textit{ybcN} locus
of the chromosome. Final strain construction was achieved by performing repeated
P1 transduction \cite{Thomason2007} of the different operator and \textit{lacI}
constructs to generate each combination used in this work. Integration was
confirmed by PCR amplification of the replaced chromosomal region and by
sequencing. Primers and final strain genotypes are listed in Appendix
K.

It is important to note that the rest of the \textit{lac} operon
(\textit{lacZYA}) was never expressed. The LacY protein is a transmembrane
protein which actively transports lactose as well as IPTG into the cell. As LacY
was never produced in our strains, we assume that the extracellular and
intracellular IPTG concentration was approximately equal due to diffusion across
the membrane into the cell as is suggested by previous work
\cite{FernandezCastane2012}.

To make this theory applicable to transcription factors with any number of DNA
binding domains, we used a different definition for repressor copy number than
has been used previously. We define the LacI copy number as the average number
of repressor dimers per cell whereas in \cite{Garcia2011}, the copy number is
defined as the average number of repressor tetramers in each cell. To motivate
this decision, we consider the fact that the LacI repressor molecule exists as a
tetramer in \textit{E. coli} \cite{Lewis1996} in which a single DNA binding
domain is formed from dimerization of LacI proteins, so that wild-type LacI
might be described as dimer of dimers. Since each dimer is allosterically
independent (i.e. either dimer can be allosterically active or inactive,
independent of the configuration of the other dimer) \cite{Daber2009}, a single
LacI tetramer can be treated as two functional repressors. Therefore, we have
simply multiplied the number of repressors reported in \cite{Garcia2011} by a
factor of two. This factor is included as a keyword argument in the numerous
Python functions used to perform this analysis, as discussed in the code
documentation.

A subset of strains in these experiments were measured using fluorescence
microscopy for validation of the flow cytometry data and results. To aid in the
high-fidelity segmentation of individual cells, the strains were modified to
constitutively express an mCherry fluorophore. This reporter was cloned into a
pZS4*1 backbone \cite{Lutz1997} in which mCherry is driven by the
\textit{lacUV5} promoter. All microscopy and flow cytometry experiments were
performed using these strains.

%%%%%%%%%%%%%%%%%%%%%%%%%%%%%%%%%%%%%%%%%%%%%%%%%%%%%%%%%%%%%%%%%%%%%%%%%%%%%%%
\subsection*{Growth Conditions for Flow Cytometry Measurements}

All measurements were performed with \textit{E. coli} cells grown to
mid-exponential phase in standard M9 minimal media (M9 5X Salts, Sigma-Aldrich
M6030; $2\,\text{mM}$ magnesium sulfate, Mallinckrodt Chemicals 6066-04; $100\,\mu\text{M}$
calcium chloride, Fisher Chemicals C79-500) supplemented with 0.5\% (w/v)
glucose. Briefly, $500\,\mu\text{L}$ cultures of \textit{E. coli} were inoculated into
Lysogeny Broth (LB Miller Powder, BD Medical) from a 50\% glycerol frozen stock
(-80$^\circ$C) and were grown overnight in a $2\,\text{mL}$ 96-deep-well plate sealed with
a breathable nylon cover (Lab Pak - Nitex Nylon, Sefar America Inc. Cat. No.
241205)~with rapid agitation for proper aeration. After approximately $12$ to
$15$ hours, the cultures had reached saturation and were diluted 1000-fold into
a second $2\,\text{mL}$ 96-deep-well plate where each well contained $500\,\mu\text{L}$ of M9
minimal media supplemented with 0.5\% w/v glucose (anhydrous D-Glucose, Macron
Chemicals) and the appropriate concentration of IPTG (Isopropyl $\beta$-D-1
thiogalactopyranoside Dioxane Free, Research Products International). These were
sealed with a breathable cover and were allowed to grow for approximately eight
hours. Cells were then diluted ten-fold into a round-bottom 96-well plate
(Corning Cat. No. 3365) containing $90\,\mu\text{L}$ of M9 minimal media
supplemented with 0.5\% w/v glucose along with the corresponding IPTG
concentrations. For each IPTG concentration, a stock of 100-fold concentrated
IPTG in double distilled water was prepared and partitioned into
$100\,\mu\text{L}$ aliquots. The same parent stock was used for all experiments
described in this work.

%%%%%%%%%%%%%%%%%%%%%%%%%%%%%%%%%%%%%%%%%%%%%%%%%%%%%%%%%%%%%%%%%%%%%%%%%%%%%%%
\subsection*{Flow Cytometry}

Unless explicitly mentioned, all fold-change measurements were collected on a
Miltenyi Biotec MACSquant Analyzer 10 Flow Cytometer graciously provided by the
Pamela Bj\"{o}rkman lab at Caltech. Detailed information regarding the voltage
settings of the photo-multiplier detectors can be found in Table
S1. Prior to each day's experiments, the analyzer was
calibrated using MACSQuant Calibration Beads (Cat. No. 130-093-607) such that
day-to-day experiments would be comparable. All YFP fluorescence measurements
were collected via $488\,\text{nm}$ laser excitation coupled with a
525/$50\,\text{nm}$ emission filter. Unless otherwise specified, all
measurements were taken over the course of two to three hours using automated
sampling from a 96-well plate kept at approximately $4^\circ \, \hbox{-} \,
10^\circ$C on a MACS Chill 96 Rack (Cat. No. 130-094-459). Cells were diluted to
a final concentration of approximately $4\times 10^{4}$ cells per $\mu\text{L}$
which corresponded to a flow rate of 2,000-6,000 measurements per second, and
acquisition for each well was halted after 100,000 events were detected. Once
completed, the data were extracted and immediately processed using the following
methods.

%%%%%%%%%%%%%%%%%%%%%%%%%%%%%%%%%%%%%%%%%%%%%%%%%%%%%%%%%%%%%%%%%%%%%%%%%%%%%%%
\subsection*{Unsupervised Gating of Flow Cytometry Data}

Flow cytometry data will frequently include a number of spurious events or other
undesirable data points such as cell doublets and debris. The process of
restricting the collected data set to those data determined to be ``real'' is
commonly referred to as gating. These gates are typically drawn manually
\cite{Maecker2005} and restrict the data set to those points which display a
high degree of linear correlation between their forward-scatter (FSC) and
side-scatter (SSC). The development of unbiased and unsupervised methods of
drawing these gates is an active area of research \cite{Aghaeepour2013, Lo2008}.
For our purposes, we assume that the fluorescence level of the population should
be log-normally distributed about some mean value. With this assumption in
place, we developed a method that allows us to restrict the data used to
compute the mean fluorescence intensity of the population to the smallest
two-dimensional region of the $\log(\mathrm{FSC})$ vs. $\log(\mathrm{SSC})$
space in which 40\% of the data is found. This was performed by fitting a
bivariate Gaussian distribution and restricting the data used for calculation to
those that reside within the 40th percentile. This procedure is described in
more detail in the supplementary information as well as in a Jupyter notebook
located in this paper's
\href{https://rpgroup-pboc.github.io/mwc_induction/code/notebooks/unsupervised_gating.html}{Github repository}.

%%%%%%%%%%%%%%%%%%%%%%%%%%%%%%%%%%%%%%%%%%%%%%%%%%%%%%%%%%%%%%%%%%%%%%%%%%%%%%%
\subsection*{Experimental Determination of Fold-Change}

For each strain and IPTG concentration, the fold-change in gene expression was
calculated by taking the ratio of the population mean YFP expression in the
presence of LacI repressor to that of the population mean in the absence of LacI
repressor. However, the measured fluorescence intensity of each cell also
includes the autofluorescence contributed by the weak excitation of the myriad
protein and small molecules within the cell. To correct for this background, we
computed the fold change as
\begin{equation}
 \text{fold-change} = \frac{\langle I_{R > 0} \rangle - \langle I_\text{auto}\rangle}{\langle I_{R = 0} \rangle - \langle I_\text{auto}\rangle},
\end{equation}
where $\langle I_{R > 0}\rangle$ is the average cell YFP intensity in the
presence of repressor, $\langle I_{R = 0}\rangle$ is the average cell YFP
intensity in the absence of repressor, and $\langle I_\text{auto} \rangle$ is
the average cell autofluorescence intensity, as measured from cells that lack
the \textit{lac}-YFP construct.

%%%%%%%%%%%%%%%%%%%%%%%%%%%%%%%%%%%%%%%%%%%%%%%%%%%%%%%%%%%%%%%%%%%%%%%%%%%%%%%
\subsection*{Bayesian Parameter Estimation}

In this work, we determine the the most-likely parameter values for the inducer
dissociation constants $K_A$ and $K_I$ of the active and inactive state,
respectively, using Bayesian methods. We compute the probability distribution of
the value of each parameter given the data $D$, which by Bayes' theorem is given
by
\begin{equation}\label{bayes_theorem}
	P(K_A, K_I \mid D) = \frac{P(D \mid K_A, K_I)P(K_A, K_I)}{P(D)},
\end{equation}
where $D$ is all the data composed of independent variables (repressor copy
number $R$, repressor-DNA binding energy $\Delta\varepsilon_{RA}$, and inducer
concentration $c$) and one dependent variable (experimental fold-change). $P(D
\mid K_A, K_I)$ is the likelihood of having observed the data given the
parameter values for the dissociation constants, $P(K_A, K_I)$ contains all the
prior information on these parameters, and $P(D)$ serves as a normalization
constant, which we can ignore in our parameter estimation.
\eref[eq_fold_change_full] assumes a deterministic relationship between the
parameters and the data, so in order to construct a probabilistic relationship
as required by \eref[bayes_theorem], we assume that the experimental fold-change
for the $i^\text{th}$ datum given the parameters is of the form
\begin{equation}
\foldchange _{\exp}^{(i)} = \left( 1 + \frac{\left(1 +
\frac{c^{(i)}}{K_A}\right)^2}{\left( 1 + \frac{c^{(i)}}{K_A}\right)^2 +
e^{-\beta \Delta \varepsilon_{AI}} \left(1 + \frac{c^{(i)}}{K_I} \right)^2} \frac{R^{(i)}}{N_{NS}} e^{-\beta
\Delta \varepsilon_{RA}^{(i)}}\right)^{-1} + \epsilon^{(i)},
\label{eq_fold_change_exp}
\end{equation}
where $\epsilon^{(i)}$ represents the departure from the deterministic
theoretical prediction for the $i^\text{th}$ data point. If we assume that these
$\epsilon^{(i)}$ errors are normally distributed with mean zero and standard
deviation $\sigma$, the likelihood of the data given the parameters is of the
form
\begin{equation} \label{eq_likelihood}
P(D \vert K_A, K_I, \sigma) =
\frac{1}{(2\pi\sigma^2)^{\frac{n}{2}}}\prod\limits_{i=1}^n \exp
\left[-\frac{(\foldchange^{(i)}_{\exp} - \foldchange(K_A, K_I, R^{(i)},
	\Delta\varepsilon_{RA}^{(i)}, c^{(i)}))^2}{2\sigma^2}\right],
\end{equation}
where $\foldchange^{(i)}_{\text{exp}}$ is the experimental fold-change and
$\foldchange(\,\cdots)$ is the theoretical prediction. The product
$\prod_{i=1}^n$ captures the assumption that the $n$ data points are
independent. Note that the likelihood and prior terms now include the extra
unknown parameter $\sigma$. In applying \eref[eq_likelihood], a choice of $K_A$
and $K_I$ that provides better agreement between theoretical fold-change
predictions and experimental measurements will result in a more probable
likelihood.

Both mathematically and numerically, it is convenient to define $\tilde{k}_A =
-\log \frac{K_A}{1\,\text{M}}$ and $\tilde{k}_I = -\log \frac{K_I}{1\,\text{M}}$
and fit for these parameters on a log scale. Dissociation constants are scale
invariant, so that a change from $10\,\mu\text{M}$ to $1\,\mu\text{M}$ leads to
an equivalent increase in affinity as a change from $1\,\mu\text{M}$ to
$0.1\,\mu\text{M}$. With these definitions we assume for the prior
$P(\tilde{k}_A, \tilde{k}_I, \sigma)$ that all three parameters are independent.
In addition, we assume a uniform distribution for $\tilde{k}_A$ and
$\tilde{k}_I$ and a Jeffreys prior \cite{Sivia2006} for the scale parameter
$\sigma$. This yields the complete prior
\begin{equation}
P(\tilde{k}_A, \tilde{k}_I, \sigma) \equiv \frac{1}{(\tilde{k}_A^{\max} -
\tilde{k}_A^{\min})} \frac{1}{(\tilde{k}_I^{\max} -
\tilde{k}_I^{\min})}\frac{1}{\sigma}.
\end{equation}
These priors are maximally uninformative meaning that they imply no prior
knowledge of the parameter values. We defined the $\tilde{k}_A$ and
$\tilde{k}_A$ ranges uniform on the range of $-7$ to $7$, although we note that
this particular choice does not affect the outcome provided the chosen range is
sufficiently wide.

Putting all these terms together we can now sample from $P(\tilde{k}_A,
\tilde{k}_I, \sigma \mid D)$ using Markov chain Monte Carlo (see
\href{https://rpgroup-pboc.github.io/mwc_induction/code/notebooks/bayesian_parameter_estimation}{GitHub repository}) to compute the most likely parameter as well as the error bars (given by the 95\% credible region) for $K_A$ and $K_I$.

%%%%%%%%%%%%%%%%%%%%%%%%%%%%%%%%%%%%%%%%%%%%%%%%%%%%%%%%%%%%%%%%%%%%%%%%%%%%%%%
\subsection*{Data Curation}

All of the data used in this work as well as all relevant code can be found at
this \href{http://rpgroup-pboc.github.io/mwc_induction}{dedicated website}. Data
were collected, stored, and preserved using the Git version control software in
combination with off-site storage and hosting website GitHub. Code used to
generate all figures and complete all processing step as and analyses are
available on the GitHub repository. Many analysis files are stored as
instructive Jupyter Notebooks. The scientific community is invited to fork our
repositories and open constructive issues on the
\href{https://www.github.com/rpgroup-pboc/mwc_induction}{GitHub repository}.

	%%%%%%%%%%%%%%%%%%%%%%%%%%%%%%%%%%%%%%%%%%%%%%%%%%%%%%%%%%%%%%%%%%%%%%%%%%%
\section*{Acknowledgements}

This work has been a wonderful exercise in scientific collaboration. We thank
Hernan Garcia for information and advice for working with these bacterial
strains, Pamela Bj\"{o}rkman and Rachel Galimidi for access and training for use
of the Miltenyi Biotec MACSQuant flow cytometer, and Colin deBakker of Milteny
Biotec for useful advice and instruction in flow cytometry. The experimental
front of this work began at the Physiology summer course at the Marine
Biological Laboratory in Woods Hole, MA operated by the University of Chicago.
We thank Simon Alamos, Nalin Ratnayeke, and Shane McInally for their work on the
project during the course. We also thank Suzannah Beeler, Justin Bois, Robert
Brewster, Ido Golding, Soichi Hirokawa, Jan\'{e} Kondev, Tom Kuhlman, Heun Jin Lee, Muir
Morrison, Nigel Orme, Alvaro Sanchez, and Julie Theriot for useful advice and discussion. We are also grateful to
the three anonymous reviewers for substantially improving the quality
of our work and our paper. This work was
supported by La Fondation Pierre-Gilles de Gennes, the Rosen Center at Caltech,
and the National Institutes of Health DP1 OD000217 (Director's Pioneer Award),
R01 GM085286, and 1R35 GM118043-01 (MIRA). Nathan Belliveau is a Howard Hughes
Medical Institute International Student Research fellow.

\subsubsection*{Competing interests} 

The authors have declared that no competing interests exist.

\subsection*{Author contributions}

MRM, SB,  NB, GC, and TE contributed equally to this work. MRM, SB, NB, GC performed experiments. TE and MRM  laid groundwork for the model. MRM, SB, NB,  GC, and TE performed the  data analysis. MRM, GC, NB, and SB wrote  code  used for all experimental analysis and parameter estimation. GC made the figures for the main text and GC, MRM, SB, and NB  made figures for the supplemental information. MRM, SB, NB, GC, TE, and RP wrote the paper.  ML and RP provided useful insight and advice in designing and executing the work.
	
	% Supporting Information files
	\include{supporting_information_title}
	
	\include{section_appendix_A_epsilonAI}
    \include{section_appendix_B_fugacity}
	\include{section_appendix_C_flow_gating}
	\include{section_appendix_D_microscopy}
	\include{section_appendix_E_sensitivity}
	\include{section_appendix_F_global_fit}
    \include{section_appendix_G_oid}
	\include{section_appendix_H_comparison}
	\include{section_appendix_I_properties}
	\include{section_appendix_J_applications}
	\include{section_appendix_K_strainlist}
	
	%%%%%%%%%%%%%%%%%%%%%% BIBLIOGRAPHY %%%%%%%%%%%%%%%%%%%%%%%%%%%%
	
	\ifx\showAll\true
		% Full bibliography
		\bibliography{library}
	\else
		\ifx\showText\true
			% Main Text Bibliography
			\setcounter{page}{24}

%\providecommand{\latin}[1]{#1}
%\providecommand*\mcitethebibliography{\thebibliography}
%\csname @ifundefined\endcsname{endmcitethebibliography}
%{\let\endmcitethebibliography\endthebibliography}{}

		\else
			% Supporting Information Bibliography
			\include{section_bibliography_SI}
		\fi
	\fi
	
\end{document}